\documentclass[aps,prc,twocolumn,superscriptaddress,floatfix]{revtex4} 
\usepackage{textcomp}
\usepackage{nicefrac}
\usepackage{longtable}
\usepackage{graphicx}
\usepackage{multirow}
\usepackage{amsmath}
\usepackage{makecell}
\usepackage{xcolor}
\usepackage{epstopdf}
\usepackage{hyperref}

\begin{document}

 \title{Isoscalar monopole and dipole transitions in $^{24}$Mg, $^{26}$Mg and $^{28}$Si}

  \author{P. Adsley}
 \email{philip.adsley@wits.ac.za}
 \affiliation{School of Physics, University of the Witwatersrand, Johannesburg 2050, South Africa}
\affiliation{Department of Physics,  Stellenbosch University, Private Bag X1, 7602 Matieland, Stellenbosch, South Africa}
 \affiliation{iThemba Laboratory for Accelerator Based Sciences, Somerset West 7129, South Africa}
 \affiliation{Institut de Physique Nucl\'{e}aire d'Orsay, UMR8608, IN2P3-CNRS, Universit\'{e} Paris Sud 11, 91406 Orsay, France}

 \author{V. O. Nesterenko}
 \affiliation{Laboratory of Theoretical Physics, Joint Institute for Nuclear Research, Dubna, Moscow Region 141980, Russia}
 \affiliation{State University \textquotedblleft Dubna\textquotedblright, Dubna, Moscow Region 141980, Russia}
 \affiliation{Moscow Institute of Physics and Technology, Dolgoprudny, Moscow Region, 141701, Russia}

 \author{M. Kimura}
 \affiliation{Department of Physics, Hokkaido University, 060-0810 Sapporo, Japan}
 \affiliation{Reaction Nuclear Data Centre, Faculty of Science, Hokkaido University, 060-0810 Sapporo, Japan}

 \author{L. M. Donaldson}
 \affiliation{iThemba Laboratory for Accelerator Based Sciences, Somerset West 7129, South Africa}
 \affiliation{School of Physics, University of the Witwatersrand, Johannesburg 2050, South Africa}

 \author{R. Neveling}
 \affiliation{iThemba Laboratory for Accelerator Based Sciences, Somerset West 7129, South Africa}

 \author{J. W. Br\"{u}mmer}
 \affiliation{Department of Physics,  Stellenbosch University, Private Bag X1, 7602 Matieland, Stellenbosch, South Africa}

 \author{D. G. Jenkins}
 \affiliation{Department of Physics, University of York, Heslington, York, YO10 5DD, United Kingdom}

  \author{N. Y. Kheswa}
 \affiliation{iThemba Laboratory for Accelerator Based Sciences, Somerset West 7129, South Africa}

 \author{J. Kvasil}
 \affiliation{Institute of Particle and Nuclear Physics, Charles University, CZ-18000, Praha 8, Czech Republic}

 \author{K. C. W. Li}
 \affiliation{Department of Physics,  Stellenbosch University, Private Bag X1, 7602 Matieland, Stellenbosch, South Africa}

 \author{D. J. Mar\'{i}n-L\'{a}mbarri}
 \affiliation{iThemba Laboratory for Accelerator Based Sciences, Somerset West 7129, South Africa}
 \affiliation{Department of Physics, University of the Western Cape, P/B X17, Bellville 7535, South Africa}
\affiliation{Instituto de F\'{i}sica, Universidad Nacional Aut\'{o}noma de M\'{e}xico, Apartado Postal 20-364, 01000 Cd. M\'{e}xico, M\'{e}xico}

 \author{Z. Mabika}
 \affiliation{Department of Physics, University of the Western Cape, P/B X17, Bellville 7535, South Africa}

 \author{P. Papka}
 \affiliation{Department of Physics,  Stellenbosch University, Private Bag X1, 7602 Matieland, Stellenbosch, South Africa}

 \author{L. Pellegri}
 \affiliation{iThemba Laboratory for Accelerator Based Sciences, Somerset West 7129, South Africa}
 \affiliation{School of Physics, University of the Witwatersrand, Johannesburg 2050, South Africa}

 \author{V. Pesudo}
 \affiliation{iThemba Laboratory for Accelerator Based Sciences, Somerset West 7129, South Africa}
 \affiliation{Department of Physics, University of the Western Cape, P/B X17, Bellville 7535, South Africa}
 \affiliation{Centro de Investigaciones Energéticas, Medioambientales y Tecnológicas, Madrid 28040, Spain}

 \author{B. Rebeiro}
 \affiliation{Department of Physics, University of the Western Cape, P/B X17, Bellville 7535, South Africa}

 \author{P.-G. Reinhard}
 \affiliation{Institut f\"ur Theoretische Physik II, Universit\"at Erlangen, D-91058, Erlangen, Germany}

 \author{F. D. Smit}
 \affiliation{iThemba Laboratory for Accelerator Based Sciences, Somerset West 7129, South Africa}

 \author{W. Yahia-Cherif}
 \affiliation{Universit\'{e} des Sciences et de la Technologie Houari Boumediene (USTHB), Facult\'{e} de Physique, B.P. 32 El-Alia,16111 Bab Ezzouar, Algiers, Algeria}

 \date{\today}

\begin{abstract}
\begin{description}
 \item[Background] Nuclei in the $sd$-shell demonstrate a remarkable interplay of cluster and
	    mean-field phenomena. The {$N=Z$ } nuclei, such as $^{24}$Mg and $^{28}$Si, have
	    been the focus of the theoretical study of both phenomena in the past. A variety of different cluster structures  in these nuclei are predicted, characterized by isoscalar
        dipole and monopole transitions. For example, low-energy isoscalar vortical dipole states were predicted in $^{24}$Mg.
	   
	    The  cluster and vortical mean-field phenomena can be probed
        by excitation of isoscalar monopole and
	    dipole states in scattering of isoscalar particles such as deuterons or
	    $\alpha$ particles. 
\item[Purpose]
       To investigate, both experimentally and theoretically, the isoscalar
	   dipole $IS1$ and monopole $IS0$ strengths in three essentially different light nuclei with different properties: stiff prolate $^{24}$Mg, soft prolate $^{26}$Mg and soft oblate
	   $^{28}$Si. To analyze possible manifestations of clustering and vorticity in these nuclei.
\item[Methods]
       Inelastically scattered $\alpha$ particles were momentum-analysed in the
	   K600 magnetic spectrometer at iThemba LABS, Cape Town, South Africa. The
	   scattered particles were detected in two multi-wire drift chambers and two
	   plastic scintillators placed at the focal plane of the K600. In the heoretical discussion, the Skyrme Quasiparticle Random-Phase Approximation (QRPA) and Antisymmetrized Molecular Dynamics + Generator Coordinate Method (AMD+GCM) were used.
\item[Results]
       A number of isoscalar monopole and dipole transitions were observed in the nuclei studied. Using this information, suggested structural assignments have been made for the various excited states.  $IS1$ and $IS0$ strengths obtained within QRPA and AMD+GCM are compared with the experimental data.
       The QRPAcalculations lead us to conclude that: i) the mean-field vorticity appears mainly in dipole states with $K=1$, ii) the dipole (monopole) states should have strong deformation-induced octupole (quadrupole) admixtures, and iii) that near the $\alpha$-particle threshold there should exist a collective state with $K=0$ for prolate nuclei and $K=1$ for oblate nuclei) with an impressive octupole strength. The results of the AMD+GCM calculations suggest that some observed states may have a mixed (mean-field + cluster) character or correspond
       to particular cluster configurations.
\item[Conclusion]
      A tentative correspondence between observed states and theoretical states from QRPA and AMD+GCM was established. The QRPA and AMD+GCM analysis shows that low-energy isoscalar dipole states combine cluster and mean-field properties.
      The QRPAcalculations show that the low-energy vorticity is well localized in $^{24}$Mg, fragmented in $^{26}$Mg, and absent in $^{28}$Si.
\end{description}
\end{abstract}

\maketitle

\section{Background}
Light nuclei demonstrate a remarkable interplay of cluster and mean-field
degrees of freedom, see e.g. the reviews of Refs. \cite{Oertzen_rew06,Horiuchi_rew12,Hornberger_rew12,KE_rew18}.
The exploration of this interplay is a demanding problem which is additionally complicated by the softness of these nuclei and related shape coexistence \cite{KE_rew18}. The low-energy isoscalar monopole ($IS0$) and dipole ($IS1$) states in light nuclei can serve as fingerprints of clustering \cite{Kimura_E0_24Mg,PhysRevC.95.044328} one of the basic features of light nuclei. $IS1$ states can also deliver important information on some mean-field features, such as vorticity \cite{PhysRevLett.120.182501,Ne_EPJ,PhysRevC.100.064302,PhysRevC.95.064319,10.1093/ptep/ptz049,PhysRevC.97.014303,PhysRevC.100.014301,chiba2019cluster}. Note that vortical currents do not contribute to the continuity equation and this flow represents an important (and, as yet, poorly explored) form of nuclear dynamics beyond the familiar irrotational motion, see the discussion in Refs. \cite{PhysRevC.84.034303,PhysRevC.89.024321}. Since dipole vortical excitations are mainly located near the particle-emission thresholds they can affect reactions rates of importance to nucleosynthesis. The exploration of low-energy $IS0$ and $IS1$ transitions in light nuclei can significantly improve our knowledge of cluster and vortical features of low-energy nuclear states.

Clustering in light $N=Z$ nuclei can manifest itself in low-lying $IS0$ transitions to $J^{\pi}=0^+$ states \cite{Kawabata20076,PhysRevC.95.044328,PhysRevC.95.024319}.
Recent theoretical work has suggested that $IS1$ excitations may also be used to explore
cluster configurations, {\it i.e.} the low-lying $0^+$ states caused by asymmetric clusters may have $1^-$ partner states, thus forming inversion doublets which indicate the symmetry of the cluster configuration \cite{PhysRevC.95.044328}. In $N\neq Z$ nuclei, the asymmetric clustering may result in enhanced electric dipole transitions between isoscalar states.

In addition to this clustering behaviour, mean-field structures may also exist. Individual low-lying vortical $IS1$ states were predicted within the Quasiparticle Random-Phase-Approximation (QRPA) \cite{PhysRevLett.120.182501,Ne_EPJ,PhysRevC.100.064302} and the Antisymmetrized Molecular Dynamics + Generator Coordinate Method (AMD+GCM) \cite{PhysRevC.95.064319,10.1093/ptep/ptz049,PhysRevC.97.014303, PhysRevC.100.014301,chiba2019cluster}.
These states should exist in $^{10}$Be \cite{PhysRevC.95.064319,10.1093/ptep/ptz049}, $^{12}$C \cite{PhysRevC.97.014303},
$^{16}$O \cite{PhysRevC.100.014301}, $^{20}$Ne  \cite{Ne_EPJ}, and $^{24}$Mg \cite{PhysRevLett.120.182501,Ne_EPJ,PhysRevC.100.064302,chiba2019cluster}.

Such individual low-lying vortical states can be differentiated from the neighbouring excitations and so much more easily resolved in experiment. Note that the intrinsic electric vortical flow of nucleons, though widely discussed in recent decades, is still very
poorly understood \cite{semenko1981vortex,PhysRevC.84.034303,PhysRevC.87.024305,PhysRevC.89.024321,nesterenko2016toroidal,EPJA19}.
The experimental observation and identification of vortical states remains a challenge for the modern experimentalist \cite{PhysRevC.100.064302}. In this respect,
 exploration of individual low-lying $IS1$ vortical states in light nuclei could be used as a promising guide in the experimental design. The ($e,e^\prime$) reaction has been recently suggested as a possible method of probing the vortical response of nuclei \cite{PhysRevC.100.064302}. The complementary $(\alpha,\alpha')$ reaction may
 be used to locate candidates for the $IS1$ vortical states for these ($e,e^\prime$) measurements.

The light nuclei $^{24}$Mg, $^{26}$Mg, and $^{28}$Si have
essentially different properties and thus represent a useful set for the comparative investigation of the interplay between the mean-field and cluster degrees of freedom.
These nuclei differ by $N/Z$ ratio, softness to deformation (stiff $^{24}$Mg and soft $^{28}$Si and $^{26}$Mg), and sign of deformation (prolate $^{24}$Mg and oblate $^{28}$Si). Therefore, it is interesting to compare the origin and behavior of low-lying $IS0$ and $IS1$ strengths in these nuclei, from the perspectives of clustering and vorticity. Many
investigations have been performed for each of these nuclei separately (see {\it e.g.} Refs. \cite{Kawabata20076,harakeh2001giant,Garg_rew19} for a general view and Refs. \cite{PhysRevC.93.044324,PhysRevC.60.014304,VANDERBORG197931,PhysRevLett.40.635,GUPTA2015343,PhysRevC.33.1116,PhysRevC.18.2457,PhysRevLett.34.1527,KAMERMANS197937,MORSCH1976386} ($^{24}$Mg), \cite{PhysRevC.97.045807,PhysRevC.85.065809,PhysRevC.80.055803,PhysRevC.79.037303,PhysRevC.96.055802,PhysRevC.93.055803,PhysRevLett.40.635,wuhr1974optical} ($^{26}$Mg), \cite{PhysRevC.96.055802,VANDERBORG1977405,PhysRevC.41.1417,PhysRevC.57.1134,wuhr1974optical,PhysRevC.65.034302,PhysRevC.45.2904,PhysRevLett.34.1527,KAMERMANS197937,MORSCH197615,MORSCH1976386} ($^{28}$Si)
for particular studies). We now provide comparative experimental and theoretical analyses of these nuclei.

In this paper, we report $IS0$ and $IS1$ strengths in $^{24}$Mg, $^{26}$Mg and $^{28}$Si, determined from $\alpha$-particle inelastic scattering at very forward scattering angles
(including zero degrees). The data were obtained with the K600 magnetic spectrometer at iThemba LABS (Cape Town, South Africa). The data are limited to excitation energy $E_x < 16$ MeV so as to avoid the regions dominated by giant resonances, where identification of individual states is difficult without observation of charged-particle decays.

The theoretical analysis is performed within the QRPA model for axially deformed nuclei \cite{BenRein2003,repko2015skyrme,Repko2017,PhysRevC.99.044307,Nest_PRC16} and the AMD+GCM model \cite{PhysRevC.95.064319,10.1093/ptep/ptz049,PhysRevC.97.014303,PhysRevC.100.014301,chiba2019cluster} which can take into account both axial and triaxial quadrupole deformations
and describe the evolution of the nuclear shape with excitation energy. Moreover, AMD+GCM includes the ability to describe the interplay between mean-field and cluster degrees of freedom.
Despite some overlap of QRPA and AMD+GCM, the models basically describe different information on nuclear properties. QRPA treats excited states with a mean-field approach and is therefore suitable for investigation of the nuclear vorticity. Meanwhile, AMD+GCM highlights cluster properties. Altogether, QRPA and AMD+GCM supplement one another and comparison of their results is vital for light nuclei. Our analysis mainly focuses on possible manifestations of clustering and vorticity in $IS0$ and $IS1$ states.

The paper is organized as follows. In Secs. \ref{sec:Experiment} and \ref{sec:Data}, the experimental method and data analysis are outlined. In Sec. \ref{sec:Results}, the obtained experimental results are reported. In Sec. \ref{sec:QRPA}, the experimental $IS1$ and $IS0$ strengths are compared with QRPA calculations. The vortical and irrotational characters of $IS1$ states are scrutinized. In Sec. \ref{sec:Comparison}, the experimental data are compared with AMD+GCM results. The cluster features of $IS1$ and $IS0$ states are inspected. In Sec. \ref{sec:Conclusions}, the conclusions are offered.

\begin{figure*} 
 
\includegraphics[width=\textwidth,angle=00]{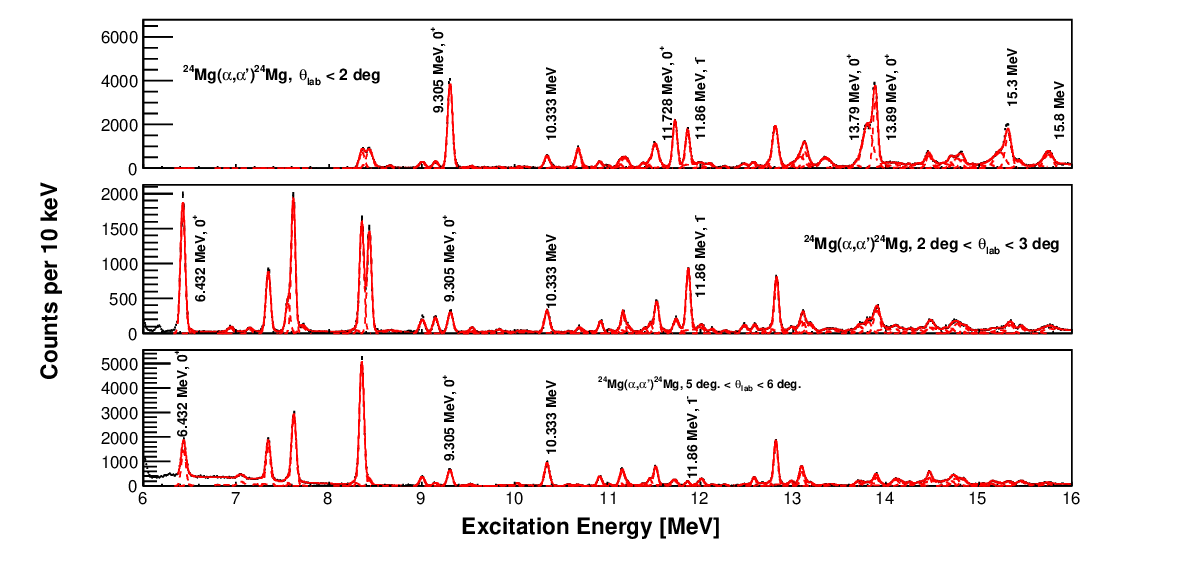}
  \caption{Fitted $^{24}$Mg spectra for: $\theta$ $<$ 2 degrees (top),
 2 $<$ $\theta$ $<$ 3 degrees (middle), and 5 $<$ $\theta$ $<$ 6 degrees
 (bottom). Some states have been labelled to guide the reader.}
 \label{fig:Mg24_fitted_spectra1}
\end{figure*}

\begin{figure*} 

\includegraphics[width=\textwidth,angle=00]{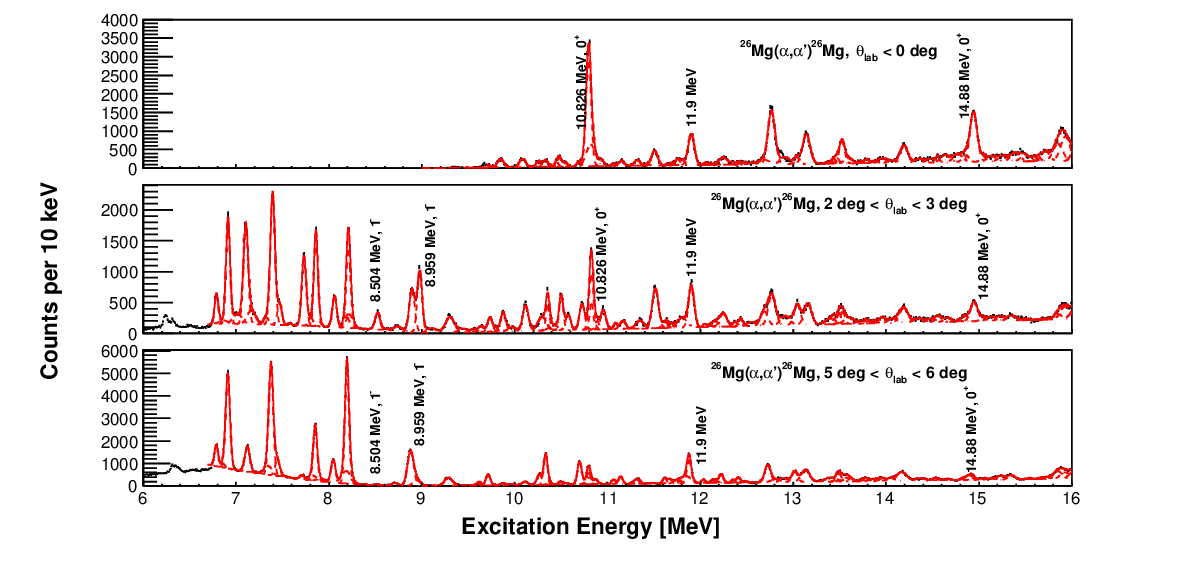}
\caption{Fitted $^{26}$Mg spectra for: $\theta$ $<$ 2 degrees (top),
 2 $<$ $\theta$ $<$ 3 degrees (middle), and 5 $<$ $\theta$ $<$ 6 degrees
 (bottom). Some states have been labelled to guide the reader. }
 \label{fig:Mg26_fitted_spectra1}
\end{figure*}

\section{Experimental Details}
\label{sec:Experiment}

A detailed description of this experiment has been given in two previous papers\cite{PhysRevC.95.024319,PhysRevC.96.055802}. A brief summary of the experimental method is given here.

A dispersion-matched beam of 200-MeV $\alpha$ particles was incident on a target and the reaction products were momentum-analysed by the K600 magnetic spectrometer. The focal-plane detectors consisted of two wire chambers giving horizontal and vertical position information, and two plastic scintillating paddles which measured energy deposited at the focal plane.

The spectrometer was used in two different modes to acquire the data: the zero-degree mode in which scattering angles of less than 2 degrees were measured, and the small-angle mode in which the spectrometer aperture covered scattering angles from 2 to 6 degrees.

For the zero-degree measurement, the background resulting from target-induced Coulomb scattering from the target necessitated running the spectrometer in a focus mode in which the scattered particles were focussed onto a vertically narrow horizontal band on the focal plane. In order to obtain a spectrum free from instrumental background a standard technique used with the iThemba K600 \cite{Neveling201129} and the RCNP Grand Raiden \cite{Tamii2009326} magnetic spectrometers was used,  in which background spectra are constructed from the regions of the focal plane above and below the focussed band. These background components are then subtracted from the signal spectrum. The vertical focussing required for this technique resulted in the loss of all vertical scattering information and limited the differential cross section for the zero-degree experiment to one point for scattering angles of less than 2 degrees.

For the small-angle measurement, the target-induced Coulomb scattering background was much lower and the spectrometer could be operated in under-focus mode, in which the vertical position on the focal plane corresponds to the vertical scattering angle into the spectrometer aperture. In this case, the scattering angle could be reconstructed from the angle with which the scattered $\alpha$ particle traversed the focal plane, and its vertical position. The angular resolution was around 0.5 degrees (FWHM) for the small-angle data. Four points were extracted for the differential cross section between 2 and 6 degrees in the laboratory frame. The procedure to calibrate the scattering angles is described in Refs. \cite{PhysRevC.95.024319,Neveling201129}.

\section{Data Analysis}
\label{sec:Data}

The techniques used for the analysis of the data have been described in more detail in Ref. \cite{PhysRevC.95.024319}. In summary, the horizontal focal-plane position was corrected for kinematic and optical aberrations according to the scattering angle into the spectrometer and the vertical focal-plane position.

The scattering angles into the spectrometer were calculated from the vertical position and the angle with which the scattered particle traverses the focal plane; these quantities were calibrated to known scattering trajectories into the spectrometer using a multi-hole collimator at the spectrometer aperture.

Horizontal focal-plane position spectra were generated for each angular region. The calibration of the focal-plane position to excitation energy used well-known states in $^{24}$Mg, $^{26}$Mg and $^{28}$Si \cite{Kawabata2013,VANDERBORG1981243}. Corrections were made according to the thickness of the relevant targets using energy losses from SRIM \cite{SRIM}.

The spectra were fitted using a number of Gaussians with a first-order polynomial used to represent background and continuum. The resolution was around 75 (65) keV (FWHM) for the zero-degree (finite-angle) data. An additional quadratic term was used at $E_x<9$ MeV for the background from $p(\alpha,\alpha)p$ elastic-scattering reactions from target contaminants. The fitted spectra for $^{24}$Mg and $^{26}$Mg at some angles are shown in Figures \ref{fig:Mg24_fitted_spectra1} and \ref{fig:Mg26_fitted_spectra1}. The $^{28}$Si spectra along with a description of the associated fitting procedures can be found in Ref. \cite{PhysRevC.95.024319}.

To quantify contamination in the targets, elastic-scattering data were taken in the small-angle mode.Population of low-lying states in nuclei contained in the target was observed. For the natural silicon target, small quantities of hydrogen, $^{12}$C, $^{16}$O and $^{29,30}$Si were observed. For the $^{24}$Mg and $^{26}$Mg targets, hydrogen, $^{12}$C and $^{16}$O were again observed but at much lower levels than for the silicon target. From previous experimental studies with the K600 (see e.g. Ref. \cite{PhysRevC.95.031302}), the locations of the $^{12}$C and $^{16}$O states are well known and excluded from further analysis.

\section{Experimental Results}
\label{sec:Results}

The focus of this paper is on the location and strength of cluster and vortical states. We report monopole ($J^\pi = 0^+$) and dipole ($J^\pi = 1^-$) states in $^{24,26}$Mg and $^{28}$Si. In addition, we discuss stateswhich have received firm or tentative monopole or dipole assignments in previous experimental studies but have not been observed in the present measurement.

The differential cross sections were extracted from the fitted spectra using:
\begin{equation}
 \frac{d\sigma}{d\Omega} = \frac{Y}{NI\eta\Delta\Omega},
\end{equation}
where $N$ is the areal density of target ions, $I$ is the integrated charge as
given by the current integrator (including the livetime fraction of the data-acquisition system), $\eta$ is the focal plane efficiency and $\Delta\Omega$
is the solid angle of the spectrometer aperture at that scattering angle. The
total efficiency, $\eta$, is the product of the efficiencies for each wire plane
per Ref. \cite{PhysRevC.95.024319}. The uncertainties in the differential cross
sections are a combination of the fitting error and Poissonian statistics.

By comparing the experimental differential cross sections to DWBA calculations
performed using the code CHUCK3 \cite{CHUCK3},
\begin{equation}
 \left( \frac{d\sigma}{d\Omega}\right)_\mathrm{exp} =
 \beta_{R,\lambda}^2 \left( \frac{d\sigma}{d\Omega}\right)_\mathrm{DWBA},
 \label{eq:betaR-factor}
\end{equation}
the transition factors  $\beta_{R,\lambda}^2$ were extracted for each dipole $(\lambda=1)$ and monopole $(\lambda=0)$  state. The
contribution  of the states to the isoscalar dipole and monopole
energy-weighted sum rules (EWSRs) were computed.
The calculations were performed in accordance with Refs.
\cite{harakeh2001giant,VANDERBORG1981243}, more details are given in Appendix \ref{app:DWBA}.

There is a systematic $\sim$20\% uncertainty due to the choice of the optical-model potentials. In the present analysis, we find that the well-known $E_x = 7.555$-MeV $J^\pi = 1^-$
state in $^{24}$Mg exhausts $2.6(5)$\% of the EWSR  which is within the expected systematic deviation when compared
with previous results of $3.1(6)$\% \cite{GUPTA2015343} and $3(1)$\% \cite{PhysRevC.60.014304}.

For some of the states contamination or background in the differential cross sections is problematic. This can occur when the level density is high e.g. around the $0^+$ states in $^{24}$Mg in the region of $E_x = 13.8-14$ MeV, where a third state lies between the two $0^+$ states, or at the minima of the differential cross section where the background is similar in size to the
cross section from the state of interest. In these cases, to avoid biasing the extracted transition strengths, a subset of points from the angular distributions has been used for comparison to the DWBA calculation.

Below, in Tables I-V and Figs. 3-4,  the monopole and dipole spectra for in $^{24}$Mg, $^{26}$Mg, and $^{28}$Si are reported. Some states are discussed in separate subsections; this is done where assignments have been updated or known states have not been observed.

\subsection{$^{24}$Mg}

\begin{figure} 
\includegraphics[width=0.4\textwidth]{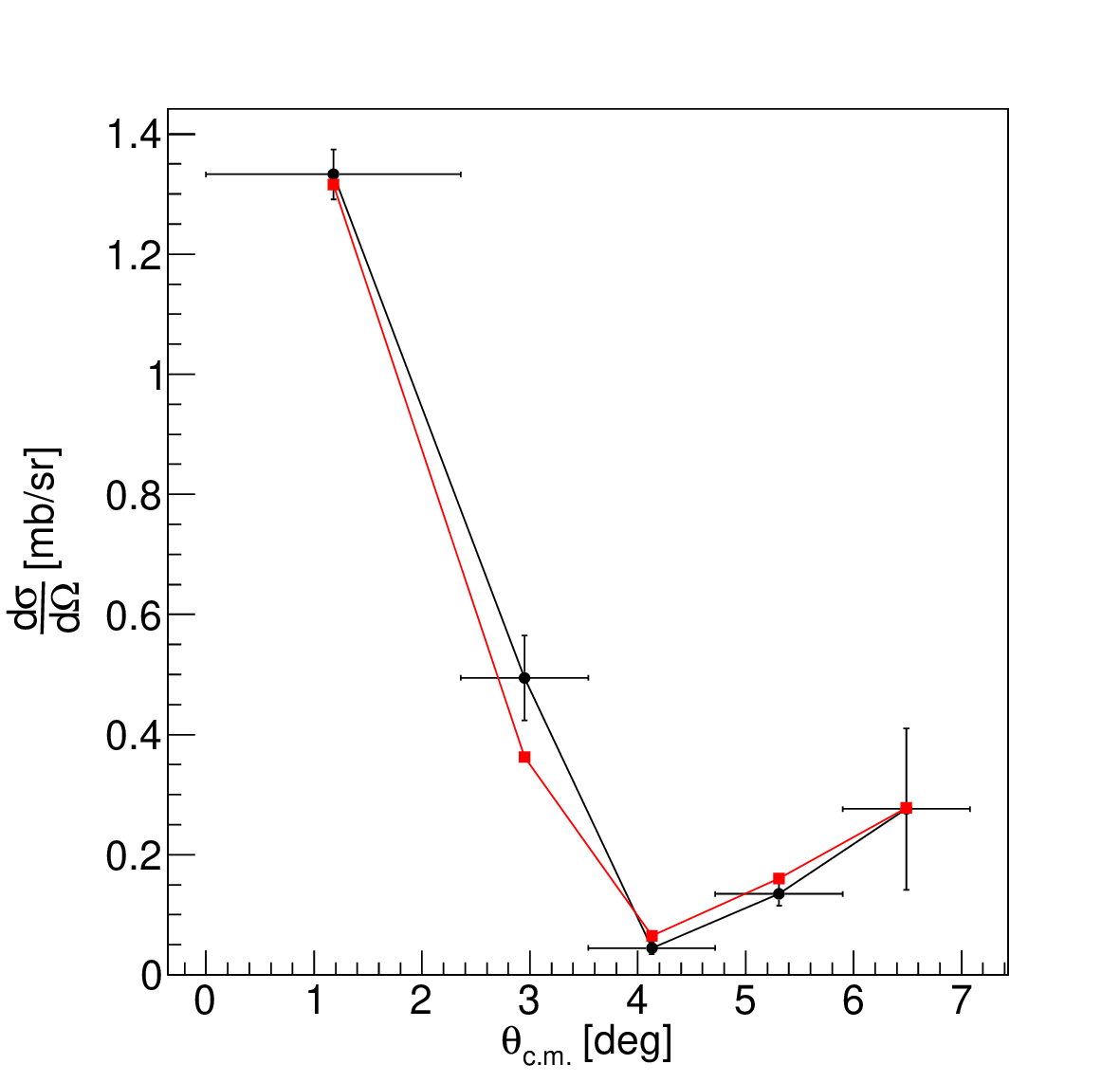}
 \caption{Differential cross section for the $J^\pi = 0^+$ state at
 10.68 MeV in $^{24}$Mg. The data are represented by points with the
 horizontal error bar delineating the angular range covered. The
 calculated angle-averaged differential cross section is shown in red.}
 \label{fig:Mg24_DCS_J_eq_0}
\end{figure}

\begin{figure} 
 \includegraphics[width=0.4\textwidth]{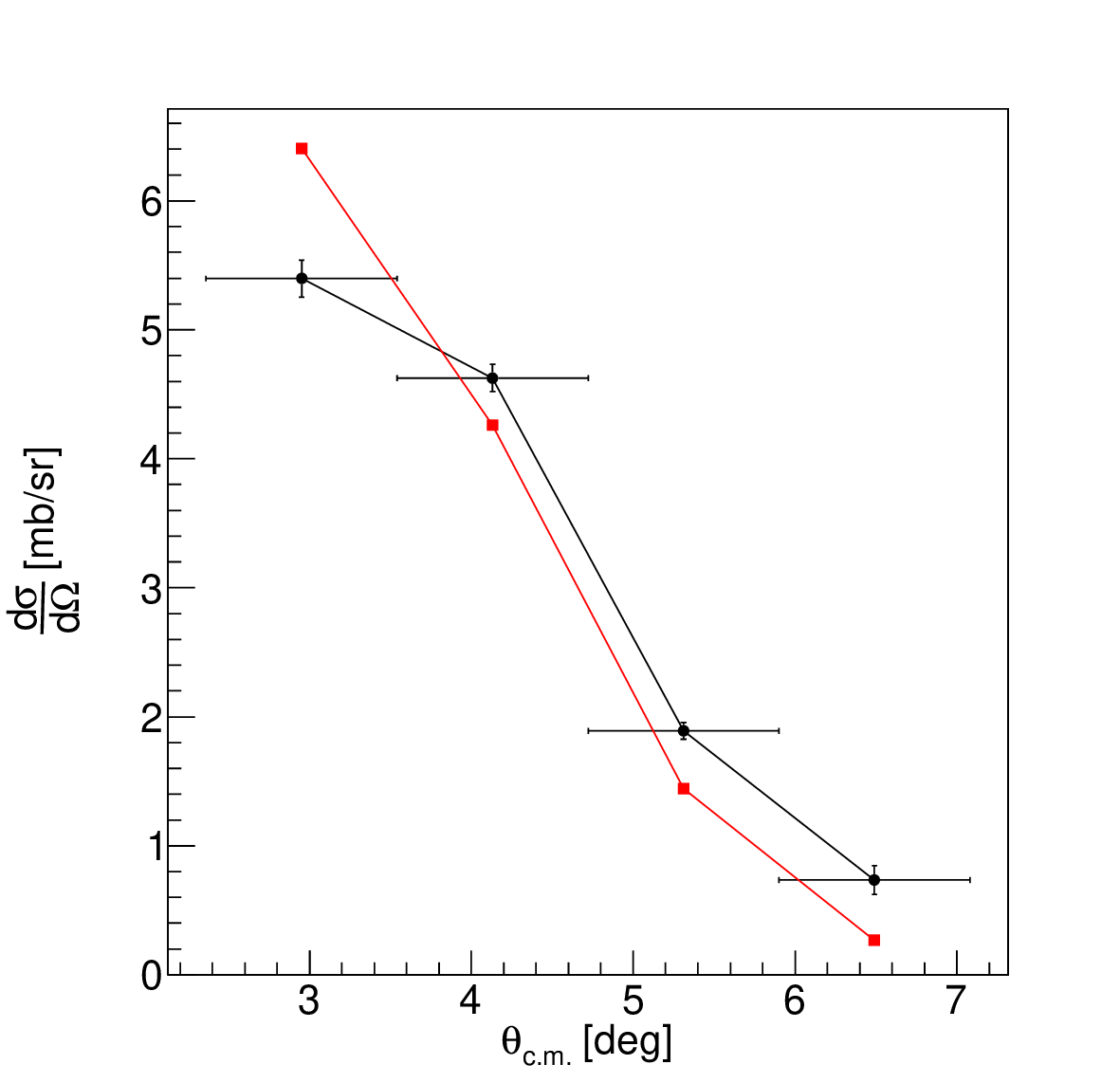}
 \caption{As Figure \ref{fig:Mg24_DCS_J_eq_0} but for the $J^\pi = 1^-$ state at 11.86 MeV in $^{24}$Mg.}
 \label{fig:Mg24_DCS_J_eq_1}
\end{figure}

A typical differential cross section for a $J^\pi = 0^+$ state in $^{24}$Mg is shown in Figure \ref{fig:Mg24_DCS_J_eq_0} and for a $J^\pi = 1^-$ state in $^{24}$Mg in Figure \ref{fig:Mg24_DCS_J_eq_1}. Similar shapes were used to identify other monopole and dipole states. The $J^\pi = 0^+$ and $J^\pi = 1^-$ levels are summarized in Tables \ref{tab:Mg24_J_eq_0_states} and \ref{tab:Mg24_J_eq_1_states}, respectively; states with the corresponding $J^\pi$ listed in the ENSDF database \cite{ENSDF} are included even when not observed.

\subsubsection{The 10.161-MeV state}

A state with $J^\pi = 0^+$ has been reported at 10.161 MeV in $^{24}$Mg($p,p^\prime$)$^{24}$Mg, $^{23}$Na($^3$He,$d$)$^{24}$Mg, $^{25}$Mg($^3$He,$^4$He)$^{24}$Mg and $^{12}$C($^{16}$O,$\alpha$)$^{24}$Mg reactions (see Ref. \cite{ENSDF} and references therein). This state is not observed in the present experiment.

\subsubsection{The 13.044/13.13-MeV state}

A $J^\pi = 0^+$ state is listed at $E_x = 13.044(3)$ MeV in Ref. \cite{ENSDF}. In the present data, a $J^\pi = 0^+$ state is observed at $E_x = 13.13$ MeV. The cause of this shift is not clea; it is possible that these are the same state and the energy has been incorrectly determined in the past or that this is an additional state.

\begin{table}
 \caption{$J^\pi = 0^+$ states in $^{24}$Mg. The excitation energies $E_x$ are, where possible, taken from Ref. \cite{ENSDF} and otherwise from the present experiment; energies are only taken from Ref. \cite{ENSDF} if a clear correspondence with a known state of the correct $J^\pi$ may be made. The  $\beta_{R,0}^2$ is the dimensionless scaling factor for the data compared to the DWBA calculations, see Appendix A for details. The $S_0$ is the percentage of the EWSR exhausted by the state.}
 \begin{ruledtabular}
 \begin{tabular}{c c c c}
$E_x$ [MeV]\footnote{From Ref. \cite{ENSDF} unless stated otherwise}
 & $\beta^2_{R,0} [10^{-4}]$ & $S_0$ & Comments \\
\hline \hline
 $6.43230(11)$ &         &         & Not on focal plane at 0\textdegree \\
 $9.30539(24)$ & $9(2)$ & $1.4(3)$ & \\
 $10.161(3)$ & & & Not observed. \\
 $10.6797(4)$ & $1.8(4)$ & $0.29(6)$ & \\
 $11.39(2)\mathrm{\footnote{Strength extracted from 0\textdegree\ data alone.}}$ & $0.6(3)$ & $0.12(2)$ & \\
 $11.7281(10)$ & $5(1)$ & $1.0(2)$ & \\
 $13.044(3)$ & & & Not observed \\
 $13.13(2)$ & $2.5(5)$ & $1.1(2)$ & \\
 $13.37(1)\mathrm{\footnote{Present experiment}}$ & $2.5(5)$ & $0.5(1)$ & \\
 $13.79(1)^{c}$ & $11(2)$ & $1.7(3)$ & \\
 $13.89(1)^{c}$ & $9(2)$ & $2.6(5)$ & \\
 $15.33(3)^{c}$ & $6(1)$ & $1.9(4)$ & \\
 $15.4364(6)$ &  & & $T=2$ \cite{ENSDF}, not observed \\
 $15.79(3)^{c}$ & $3.7(7)$ & $1.1(2)$ & \\
 \end{tabular}
 \end{ruledtabular}
\label{tab:Mg24_J_eq_0_states}
\end{table}

\begin{table}
 \caption{As in Table I but for
 $J^\pi = 1^-$ states in $^{24}$Mg.}
 \begin{ruledtabular}
 \begin{tabular}{c c c c}
 $E_x$ [MeV]\footnote{From Ref. \cite{ENSDF} unless stated otherwise} & $\beta^2_{R,1} [10^{-4}]$ & $S_1$ & Comments \\
 \hline
  $7.55504(15)$ & $0.78(16)$ & $2.6(5)$ & \\
  $8.43731(13)$ & $2.7(5)$ & $10(2)$ & \\
  $9.14599(15)$ & $0.58(12)$ & $2.4(5)$ & \\
  $11.3898(11)$ & & & Not observed \\
  $11.8649(13)$ & $2.1(4)$ & $11(2)$ & \\
  $13.19(2)\mathrm{\footnote{Present experiment}}$  & $0.49(10)$  & $2.9(6)$ &\\
 \end{tabular}
 \end{ruledtabular}
\label{tab:Mg24_J_eq_1_states}
\end{table}

\subsection{$^{26}$Mg}

Table \ref{tab:Mg26_J_eq_0_states} summarizes known $J^\pi = 0^+$ states in $^{26}$Mg either listed in the ENSDF database \cite{ENSDF} or observed during the present experiment.

\begin{table}
\caption{As in Table I but for $J^\pi = 0^+$ states in $^{26}$Mg.}
\begin{ruledtabular}
 \begin{tabular}{c c c c}
 $E_x$ [MeV]\footnote{From Ref. \cite{ENSDF} unless stated otherwise} & $\beta^2_{R,0} [10^{-4}]$ & $S_0$ & Comments \\
 \hline \hline
 $4.97230(13)$ & & & Not on focal plane at 0\textdegree \\
 $6.2562(14)$ & & & Not on focal plane at 0\textdegree \\
 $7.200(20)$ &  & & \makecell{Not on focal plane at 0\textdegree \\ Not observed\\$J^\pi = (0,1)^+$ \cite{ENSDF}} \\
 $7.428(3)$ & & & \makecell{Not on focal plane at 0\textdegree \\ Not observed\\$J^\pi = (0,1)^+$ \cite{ENSDF}} \\
 $10.159(3)$ & & & Not observed \\
 $10.74(2)$\footnote{From Ref. \cite{VANDERBORG1981243}} & & & Not observed or $J^\pi \neq 0^+$ \\
 $10.818(1)$\footnote{See Ref. \cite{adsley2020re} for a discussion of the energy of this level.} & $7(1)$ & $1.0(2)$ & \makecell{Part of a multiplet:\\ see text, Refs. \cite{PhysRevC.96.055802,PhysRevC.97.045807,adsley2020re}.} \\
 $12.345(2)$ & & & \makecell{Not observed\\ $J=0$, parity unknown} \\
 $12.72(2)^{\mathrm{c}}$ & $6(1)$ & $0.9(2)$ & \\
 $13.1(2)^c$ & $1.6(3)$ & $0.28(6)$ & New \\
 $13.5(2)^c$ & $2.0(4)$ & $0.37(7)$ & New \\
 $14.88(2)$\footnote{Present experiment} & $7(1)$ & $1.4(3)$ & \\
 \end{tabular}
 \end{ruledtabular}
 \label{tab:Mg26_J_eq_0_states}
\end{table}

Table \ref{tab:Mg26_J_eq_1_states} summarizes known $J^\pi=1^-$ states along with electrical $B(E1)$s from Refs. \cite{PhysRevC.79.037303} and \cite{PhysRevC.82.025802}.
For the data of Ref. \cite{PhysRevC.82.025802}, the partial widths of the ground-state decay are given and are converted to the reduced matrix element using the relation:

\begin{equation}
 \Gamma(\lambda \ell) = \frac{8\pi (\ell+1)}{\ell[(2\ell+1)!!]^2}
 \left(\frac{E_\gamma}{\hbar c}\right)^{2\ell+1} B(\lambda \ell)
\end{equation}
for a radiation of multipolarity $\ell$ and type (electric/magnetic) $\lambda$.
$E_\gamma$ is the energy of the $\gamma$-ray transition.

\begin{table*}[htbp]
\caption{The same as in Table I but for $J^\pi = 1^-$ states in $^{26}$Mg. Electric dipole reduced transition probabilities $B(E1)$ from Refs. \cite{PhysRevC.79.037303} and \cite{PhysRevC.82.025802} are also shown.}
\begin{ruledtabular}
 \begin{tabular}{c c c c c c}
 $E_x$ [MeV]\footnote{From Ref. \cite{ENSDF} unless stated otherwise} & $\beta^2_{R,1} [10^{-4}]$ & $S_1$ & \makecell{$B(E1,0_{gs}^+ \rightarrow 1^-)$\\ $(10^{-4} e^2\ \mathrm{fm}^2)$ \cite{PhysRevC.79.037303}}  & \makecell{$B(E1,0_{gs}^+ \rightarrow 1^-)$ \\ $(10^{-4} e^2\ \mathrm{fm}^2)$ \cite{PhysRevC.82.025802}}  &
 Comments \\
 \hline \hline
 $7.06190(20)$ & $0.34(6)$ & $1.1(2)$ & & \\
 $7.6968(8)$ & $0.76(15)$ & $2.6(5)$ & $9.4(31)$ & &  \\
 $8.5037(3)$ & $0.21(4)$ & $0.8(2)$ & $33.3(41)$ & &  \\
 $8.9594(5)$ & $0.95(19)$ & $3.9(8)$ & $12.5(22)$ & &  \\
 $9.1395(13)$ & & & $0.17(4)$ & &  \makecell{Not observed.\\Parity uncertain \cite{PhysRevC.79.037303}}\\
 $9.7708(9)$ & & & $0.58(14)$ & &  \makecell{Not observed.\\Parity is tentatively negative \cite{PhysRevC.79.037303}}\\
 $9.87(2)^b$ & $0.23(5)$ & $1.0(2)$ & & & New \\
 $10.1031(7)$ & $0.46(9)$ & $2.1(4)$ & $18.9(33)$ & &  \\
 $10.50(2)$ & $0.48(10)$ & $2.3(5)$ & & \\
 $10.5733(8)$ & $0.26(5)$ & $1.3(3)$ & & $0.75(19)$ &  \\
 $10.8057(7)$ & & & & $1.2(3)$ & \makecell{Not cleanly observed due\\ to $10.826$-MeV $J^\pi = 0^+$ state} \\
 $10.9491(8)$ & $0.29(6)$ & $1.5(3)$ & & $2.71(42)$ &  \\
 $11.28558(5)$ & & & & & From $^{25}$Mg$+n$ \cite{PhysRevC.85.044615} \\
 $11.32827(5)$ & & & & & From $^{25}$Mg$+n$ \cite{PhysRevC.85.044615} \\
 $11.51(2)$\footnote{Present experiment} & $0.67(13)$ & $3.5(7)$ & & &  \makecell{Possible multiplet \\ see Ref. \cite{PhysRevC.96.055802}}\\
 \end{tabular}
 \end{ruledtabular}
 \label{tab:Mg26_J_eq_1_states}
\end{table*}

\subsubsection{The 7.062-MeV state}

This state is listed in ENSDF \cite{ENSDF} but not observed in a previous $^{26}$Mg($\alpha,\alpha^\prime$)$^{26}$Mg reaction of Ref. \cite{VANDERBORG1981243}. In the present experiment, a state is observed at $E_x = 7.10$ MeV with a differential cross section that is consistent with a $J^\pi = 1^-$ assignment.

\subsubsection{The 10.159-MeV state}

The $J^\pi = 0^+$ state at $E_x = 10.159$ MeV in $^{26}$Mg listed in Ref. \cite{ENSDF} is not observed in the present experiment. The state has been previously observed in $^{24}$Mg($t,p$)$^{26}$Mg with $\ell=0$ \cite{ALFORD1986189} and in $^{26}$Mg($p,p^\prime$)$^{26}$Mg (see Ref. \cite{ENSDF} and references therein). We assume that the state has $T=1$ if it is populated in $^{24}$Mg($t,p$)$^{26}$Mg reactions. Therefore, population of this state in $^{26}$Mg($\alpha,\alpha^\prime$)$^{26}$Mg is unlikely to be isospin-forbidden. The reason why this state is not populated remains unclear.

\subsubsection{The 10.74-MeV tate}

Ref. \cite{VANDERBORG1981243} lists a tentative $J^\pi = 0^+$ state at $E_x = 10.74(2)$ MeV. In the present experiment, a state is observed at around $E_x = 10.72(2)$ MeV but the differential cross section is consistent with $J\geq2$.

\subsubsection{States in the region of 10.80 to 10.83 MeV}

A state with $J^\pi = 1^-$ has been identified at 10.805 MeV in $^{26}$Mg($\gamma,\gamma^\prime$)$^{26}$Mg experiments \cite{PhysRevC.80.055803}. In a preceding paper focussing on a narrow subset of astrophysically important states in $^{26}$Mg, we demonstrated that the strong state observed in the $^{26}$Mg($\alpha,\alpha^\prime$)$^{26}$Mg reaction has $J^\pi = 0^+$ and is, therefore, evidently a different state from the $J^\pi = 1^-$ state \cite{PhysRevC.96.055802}. The existence of multiple states was confirmed by a high-resolution experiment using the Munich Q3D \cite{PhysRevC.97.045807}.

In the present case, the extraction of the dipole strength is hindered by the close proximity of the strong $J^\pi = 0^+$ state. A higher-resolution inclusive measurement or a coincidence measurement of $^{26}$Mg($\alpha,\alpha^\prime\gamma$)$^{26}$Mg is necessary for the extraction of the isoscalar dipole transition strength for this state.

\subsubsection{The $11.321$-MeV state}

Notably, one $\alpha$-particle cluster state in $^{26}$Mg has been identified through direct reactions. The resonance at $E_\alpha = 0.83$ MeV observed in $^{22}$Ne($\alpha,\gamma$)$^{26}$Mg \cite{PhysRevC.99.045804,Wolke1989} and $^{22}$Ne($\alpha,n$)$^{25}$Mg \cite{GIESEN199395,1993ApJ...414..735D,PhysRevLett.87.202501} reactions clearly has a $^{22}$Ne$+\alpha$ cluster structure. However, the spin and parity of this state were not clearly assigned in previous $^{26}$Mg($\alpha,\alpha^\prime$)$^{26}$Mg reactions including our prior publication \cite{PhysRevC.96.055802,PhysRevC.93.055803}. Based on direct measurements of the resonance strengths and the inferred $\alpha$-particle width, the state almost certainly has $J^\pi = 0^+$ or $J^\pi = 1^-$ \cite{JAYATISSA2020135267,OTA2020135256}.

We do not observe any strong candidate for this state in our present experimental work and, therefore, cannot provide a monopole or dipole transition for the state.

\subsubsection{The $11.289$- and $11.329$-MeV states}

Both of these states have been identified as $J^\pi = 1^-$ using the reactions of neutrons with $^{25}$Mg. While $\gamma$-ray partial widths are available, the branching of these states is not, and, therefore, the $B(E1)$ for the ground-state transition cannot be determined.

\subsubsection{The 12.345-MeV State}

A state is listed in Ref. \cite{ENSDF} as having $J=0$ with unknown parity and $\Gamma = 40(5)$ keV. This state is not observed in the present experiment.

\subsection{$^{28}$Si}

Data on the states observed in $^{28}$Si have been previously reported in Ref. \cite{PhysRevC.95.024319}. In the present paper, we have extended the analysis up to 16 MeV to cover the same range as for the magnesium isotopes. Additional $J^\pi = 0^+$ states are observed at 15.02 and 15.76 MeV. A number of $J^\pi = 1^-$ states have been observed.

The natural silicon target contains some carbon and oxygen contamination. Carbon and oxygen states which are strongly populated in $\alpha$-particle inelastic scattering at $E_\alpha = 200$ MeV are known from previous studies with the K600 \cite{PhysRevC.95.031302} and are excluded from the reported states.

\begin{table}
\caption{As in Table I but for $J^\pi = 0^+$ states in $^{28}$Si.}
\begin{ruledtabular}
 \begin{tabular}{c c c c}
 $E_x$ [MeV]\footnote{From Ref. \cite{ENSDF} unless stated otherwise} & $\beta^2_{R,0} [10^{-4}]$ & $S_0$ & Comments \\
 \hline
 $4.97992(8)$& & &Not on focal plane at 0\textdegree \\
 $6.69074(15)$ & & & Not on focal plane at 0\textdegree \\
 $9.71(2)$\footnote{From Ref. \cite{PhysRevC.95.024319}} & $2.6(5)$ & $0.38(8)$ & \\
 $10.81(3)^b$ & $2.2(4)$ & $0.35(7)$ & \\
 $11.142(1)$\footnote{From Ref. \cite{adsley2019re}} & $5.5(11)$ & $0.9(2)$ & \makecell{See Refs. \cite{PhysRevC.95.024319,adsley2019re}.}\\
 $12.99(2)^b$ & $4.3(9)$ & $0.8(2)$ & \makecell{Unresolved multiplet. \cite{ENSDF,PhysRevC.95.024319}.} \\
 $15.02(3)^d$ & $1.4(3)$ & $0.8(2)$ & Newly observed. \\
 $15.73(3)$\footnote{Present experiment} & $2.3(5)$ & $0.32(6)$ & \makecell{May correspond to a tentative \\$E_x = 15.65(5)$-MeV \\$J^\pi = 0^+$ state \cite{VANDERBORG1981243}.} \\
 \end{tabular}
 \end{ruledtabular}
 \label{tab:Si28_J_eq_0_states}
\end{table}

\begin{table}
 \caption{As in Table I but for $J^\pi = 1^-$ states in $^{28}$Si.}
 \begin{ruledtabular}
 \begin{tabular}{c c c c}
 $E_x$ [MeV]\footnote{From Ref. \cite{ENSDF} unless stated otherwise}
 & $\beta^2_{R,1} [10^{-4}]$ & $S_1$ & Comments \\
 \hline
 $8.9048(4)$ & $1.1(2)$ & $4.8(9)$ & \\
 $9.929(17)$ & $2.3(5)$ & $11(2)$ & \makecell{Confirms a tentative \\$J^\pi = 1^-$ assignment \cite{VANDERBORG1981243}.}\\
 $10.994(2)$ & $1.2(2)$ & $6(1)$ & Ref. \cite{ENSDF} gives $J^\pi = (1,2^+)$  \\
 $11.2956(2)$ & $0.47(10)$ & $2.5(5)$ & \makecell{Confirms a tentative \\$J^\pi = 1^-$ assignment \cite{VANDERBORG1981243}.} \\
 $11.58(2)$\footnote{Present experiment} & $0.17(3)$ & $0.9(2)$ & \\
 $13.95(2)^{b}$ & $0.59(12)$ & $3.8(8)$ & \\
 \end{tabular}
  \end{ruledtabular}
\label{tab:Si28_J_eq_1_states}
\end{table}

\subsubsection{The 11.142- and 11.148-MeV states}

As explained in the previous K600 paper on $^{28}$Si($\alpha,\alpha^\prime$)$^{28}$Si, the literature lists two unresolved $J^\pi = 0^+$ and $J^\pi = 2^+$ states at 11.141 and 11.148 MeV, respectively \cite{PhysRevC.95.024319}. Further investigation of the existing data on $^{28}$Si \cite{adsley2019re} has showed that there is, in fact, only one state with $J^\pi = 0^+$ at this energy and so it is not necessary to include contributions from two states.

\subsubsection{The 11.65-MeV state}

Ref. \cite{VANDERBORG1981243} reports a tentative $J^\pi = 1^-$ state at $E_x = 11.65(2)$ MeV corresponding to a state at $E_x = 11.671$ MeV. This state is not observed in the present experiment.

\section{Comparison with QRPA calculations}
\label{sec:QRPA}

\subsection{Calculation Scheme}

We use a fully self-consistent QRPA approach \cite{repko2015skyrme,Repko2017} with
the Skyrme force SLy6 \cite{SLy6}. This force was found to be optimal in the previous calculations of dipole excitations in medium-heavy nuclei \cite{PhysRevLett.120.182501,PhysRevC.78.044313}. The nuclear mean field is computed by the code SKYAX \cite{SKYAX} using a two-dimensional mesh in cylindrical coordinates. The mesh spacing is 0.7 fm. The calculation box extends up to 3 nuclear radii. The equilibrium deformation of nuclei is obtained by minimization of the nuclear energy. The volume pairing is treated with the Bardeen-Cooper-Schrieffer (BCS)
 method \cite{Repko2017}. The pairing was found to be weak (with a pairing gap about 1 MeV) in all the cases with the exception of the neutron system in $^{26}$Mg. The QRPA is implemented in the matrix form \cite{repko2015skyrme}. The particle-hole (1ph) configuration space extends up to 80 MeV, which allows the calculations to exhaust the isoscalar $E0$ and $E1$ energy-weighted sum rules  \cite{harakeh2001giant}. The center-of-mass and pairing-induced spurious admixtures are extracted following the prescription of Ref. \cite{PhysRevC.99.044307}.

The obtained axial quadrupole deformations are $\beta_2=0.536, 0.355$ and $-0.354$ for  $^{24,26}$Mg and $^{28}$Si, respectively, meaning that $^{24,26}$Mg are taken to be prolate nuclei while $^{28}$Si is treated as oblate. In the SLy6 calculations, $^{26}$Mg has comparable oblate and prolate energy minima. Following the experimental data of Stone \cite{Stone} as well as AMD+GCM \cite{watanabe2014ground} and Skyrme \cite{PhysRevC.86.024614} calculations, the ground-state deformation of $^{26}$Mg is prolate and we use the equilibrium deformation $\beta_2$ = 0.355 from the prolate minimum for $^{26}$Mg.

Note that the absolute values obtained for equilibrium deformations are smaller than the experimental ones ($\beta^{\rm exp}_2=0.613, 0.484, -0.412$ for $^{24,26}$Mg, $^{28}$Si) \cite{bnl}. This is a common situation for deformation-soft nuclei. Indeed, $\beta^{\rm exp}_2$ are obtained from the $B(E2)$ values for the transitions in the ground-state rotational bands.
However, in soft nuclei, $B(E2)$ values include large dynamical correlations and so this leads to overestimation of the magnitude of the quadrupole deformation, $|\beta_2|$. Therefore, the present observation that $|\beta_2| < |\beta_2^{\rm exp}|$ is reasonable.

The isoscalar reduced transition probabilities
\begin{equation}
B(IS\lambda\mu)_{\nu}=|\langle  \nu |M(IS\lambda\mu)| 0 \rangle|^2,
\end{equation}
for the transitions from the ground state $|0\rangle$ with $I^{\pi}K=0^{+}0_{gs}$ to the excited $\nu$-th QRPA state with $I^{\pi}K=\lambda^{\pi}\mu$ are calculated using the monopole $IS0$ and dipole $IS1K$ transition operators:
\begin{eqnarray}
\label{M(IS0)}
\hat{M}(IS0)&=& \sum_{i=1}^A (r^2 Y_{00})_i,
\\
\hat{M}(IS1K)&=& \sum_{i=1}^A (r^3 Y_{1K})_i, \; K = 0,1
\label{M(IS1)}
\end{eqnarray}
where $Y_{00}=1/\sqrt{4\pi}$. To investigate the deformation-induced monopole/quadrupole and dipole/octupole mixing, we also compute quadrupole $B(IS20)$ and octupole $B(IS3K)$ transition probabilities
for isoscalar transitions $0^{+}0_{gs} \to 2^+0_{\nu}$ and
$0^{+}0_{gs} \to 3^-K_{\nu}$ using  transition operators
\begin{eqnarray}
\label{M(IS20)}
\hat{M}(IS20)&=& \sum_{i=1}^A (r^2 Y_{20})_i,
\\
\hat{M}(IS3K)&=& \sum_{i=1}^A (r^3 Y_{3K})_i, \; K = 0,1 .
\label{M(IS3K)}
\end{eqnarray}

We now consider the vortical and compression isoscalar strengths, $B(IS1Kv)_{\nu}$ and $B(IS1Kc)_{\nu}$, using current-dependent operators from Refs. \cite{PhysRevLett.120.182501,PhysRevC.100.064302}. We need these strengths to estimate the relative vortical and irrorational compression contributions to the dipole states. The current-dependent compression operator includes divergence of the nuclear current and so can be reduced to Eq. (\ref{M(IS1)}) using the continuity equation. For the sake of simplicity, we will further omit the dependence on $\nu$
in rate notations.

\subsection{$IS1$ strength distributions}

\begin{figure} 
\label{fig:24Mg_S_QRPA}
\includegraphics[width=8.5cm]{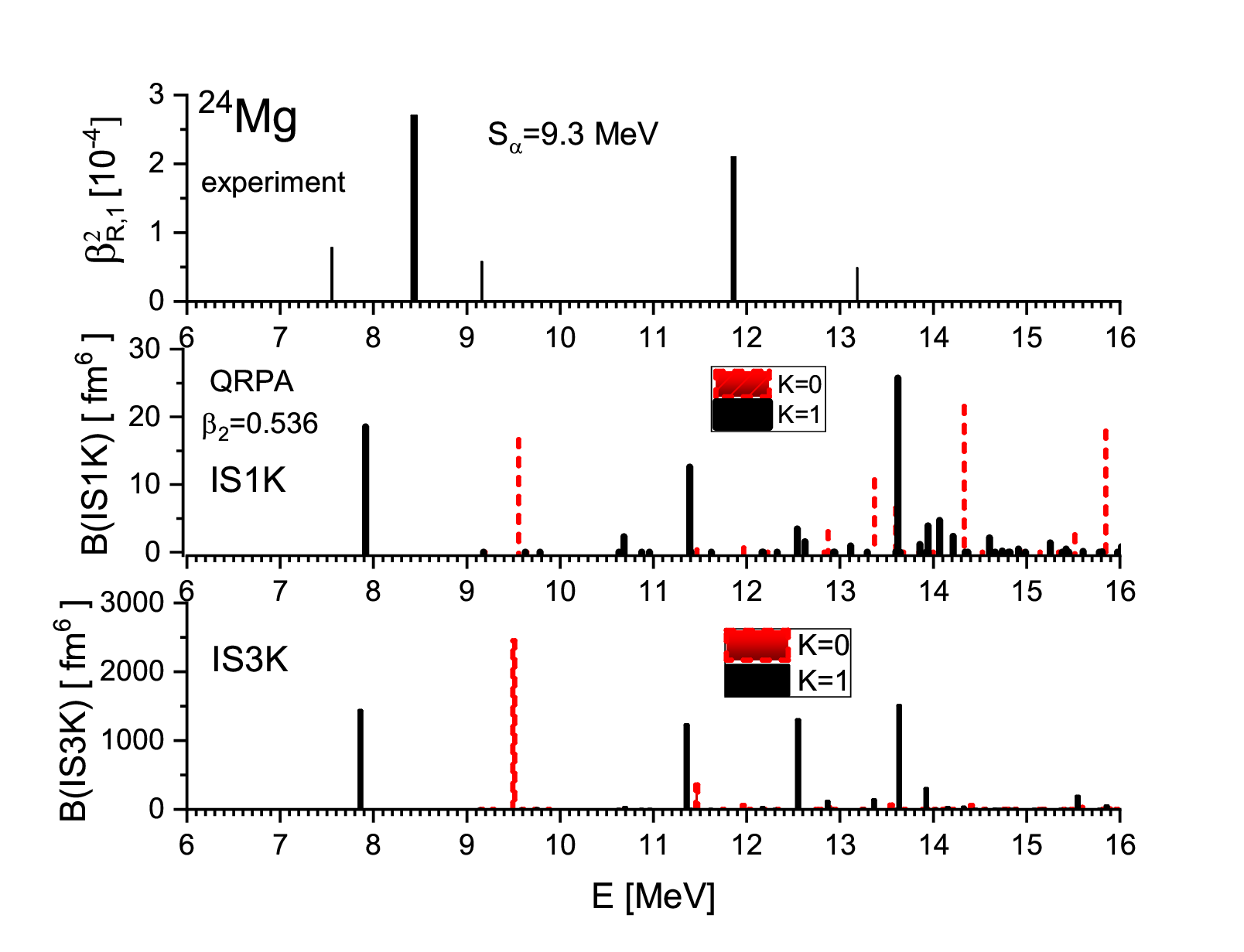}
\caption{Experimental transition factors $\beta^2_{R,1}$ (upper), QRPA isoscalar dipole compression strength, $B(IS1K)$ for $K=0$ and $K=1$ (middle), and isoscalar octupole strength $B(IS3K)$ (bottom) in $^{24}$Mg. The $\alpha$-particle threshold energy $S_{\alpha}$ and QRPA equilibrium deformation $\beta_2$ are displayed.}
\end{figure}

\subsubsection{$^{24}$Mg}

In Fig. 5, the $(\alpha,\alpha')$ experimental data (transition factors $\beta^2_{R,1}$) for $^{24}$Mg (upper plot) are compared with $B(IS1K)$ values (middle plot) for QRPA states with $K=0$ (red) and $K=1$ (black). We see that experiment and QRPA give the lowest dipole states at a similar energy, 7.56 and 7.92 MeV, respectively. In QRPA, the states 7.92-MeV ($K=1$) and
9.56-MeV ($K=0$) have large $B(IS1)$ responses and so should be well populated in the $(\alpha,\alpha')$ reaction. However, it is still difficult to establish one-to-one correspondence between these QRPA states and observed excitations, see the discussion in Ref. \cite{PhysRevLett.120.182501}. In general, QRPA gives many more dipole states between $E_x=7-16$~MeV than the observed spectrum. The calculated summed $B(IS1)$ strength is given
in Table \ref{tab-1}.

 The bottom panel of Figure 5 shows QRPA $B(IS3K)$ strengths for isoscalar octupole transitions $0^{+}0_{gs} \to 3^-K_{\nu}$. The dipole $1^-K_{\nu}$ and octupole $3^-K_{\nu}$ states
belong to the same rotational band built on the band-head state $|\nu\rangle$. Thus, the $B(IS3K)_{\nu}$ represents the level of deformation-induced octupole correlations in the band-head $|\nu\rangle$. We see that the lowest states 7.92-MeV ($K=1$) and 9.56-MeV ($K=0$) exhibit fundamental octupole strengths: $B(IS31)=715$~fm$^6$ (21 W.u.) and $B(IS30)=2450$~fm$^6$ (72 W.u.), respectively.
 Such large $B(IS3K)$ values originate from two sources: i) collectivity of the states and ii) that the dominant proton and neutron $1ph$ components of the states ($pp[211\uparrow - 330\uparrow]$, $nn[211\uparrow - 330\uparrow]$ for 7.92-MeV K=1 state
 and  $pp[211\downarrow - 101\downarrow]$, $nn[211\downarrow - 101\downarrow]$
 for 9.56-MeV K=0
 state) fulfill the selection rules for $E3K$ transitions \cite{Ni65}:
\begin{eqnarray}
\Delta K&=&0: \;  \Delta N = \pm 1,\pm 3, \;  \Delta n_z = \pm1,\pm 3, \; \Delta \Lambda = 0,
\nonumber
\\
\Delta K&=&1: \;  \Delta N = \pm 1,\pm 3, \;  \Delta n_z = 0,\pm 2, \; \Delta \Lambda = 1 .
\nonumber
\end{eqnarray}
Here, the single-particle states are specified by Nilsson asymptotic quantum numbers $N n_z \Lambda$ \cite{MN1959}, whilst the arrows indicate spin direction.
The large  $B(IS3K)$ values signify that 7.92-MeV K=1 and 9.56-MeV K=0
 states are of a mixed octupole-dipole character. Their leading $1ph$ components correspond to $\Delta N = 1$ transitions between the valence and upper quantum shells, so these states  can belong to the Low-Energy Octupole Resonance (LEOR) \cite{harakeh2001giant,Malov76}.

 As may be seen in Figure 5, both $IS1K$ and $IS3K$ distributions can be roughly separated into two groups, the first located below (7-10 MeV)  and the second located above (11-14 MeV) the $\alpha$-particle threshold ($S_{\alpha} = 9.3$~MeV).  Moreover, at the energy close to $S_{\alpha}$, there is a $E_x = 9.56$-MeV $K=0$ state with a huge $B(IS30)$ strength, which perhaps signals the octupole-deformation softness of the nucleus at this energy. It is reasonable to treat the states below $S_{\alpha}$ as being of mean-field origin, while the states close to and above $S_{\alpha}$ (including the $E_x = 9.56$-MeV $K=0$ near-threshold state) as those including cluster degrees of freedom. This is confirmed  by recent  AMD+GCM calculations for $^{24}$Mg \cite{chiba2019cluster}, where similar results  were obtained: the lowest mean-field 9.2-MeV $K=1$ state is of mean-field character and the $E_x = 11.1$-MeV state has cluster properties.

\begin{table} 
\caption{QRPA isoscalar $B(IS1)$ compression strength (in fm$^6$) summed
at the energy interval 0-16 MeV.}
\label{tab-1}
\begin{ruledtabular}
\begin{tabular}{c | c c c}
 Nucleus  & \multicolumn{3}{c}{QRPA} \\
\hline
  & $B(IS1, K=0)$ & $B(IS1, K=1)$ & $B(IS1, \mathrm{total})$
\\
\hline
\\
$^{24}$Mg & 80  & 82 & 162
\\
$^{26}$Mg & 90  & 141 & 230
\\
$^{28}$Si &  21  & 168 & 189
\\
\end{tabular}
\end{ruledtabular}
\\
\end{table}

\begin{figure} 
\label{fig:24Mg_vc}
\includegraphics[width=8.5cm]{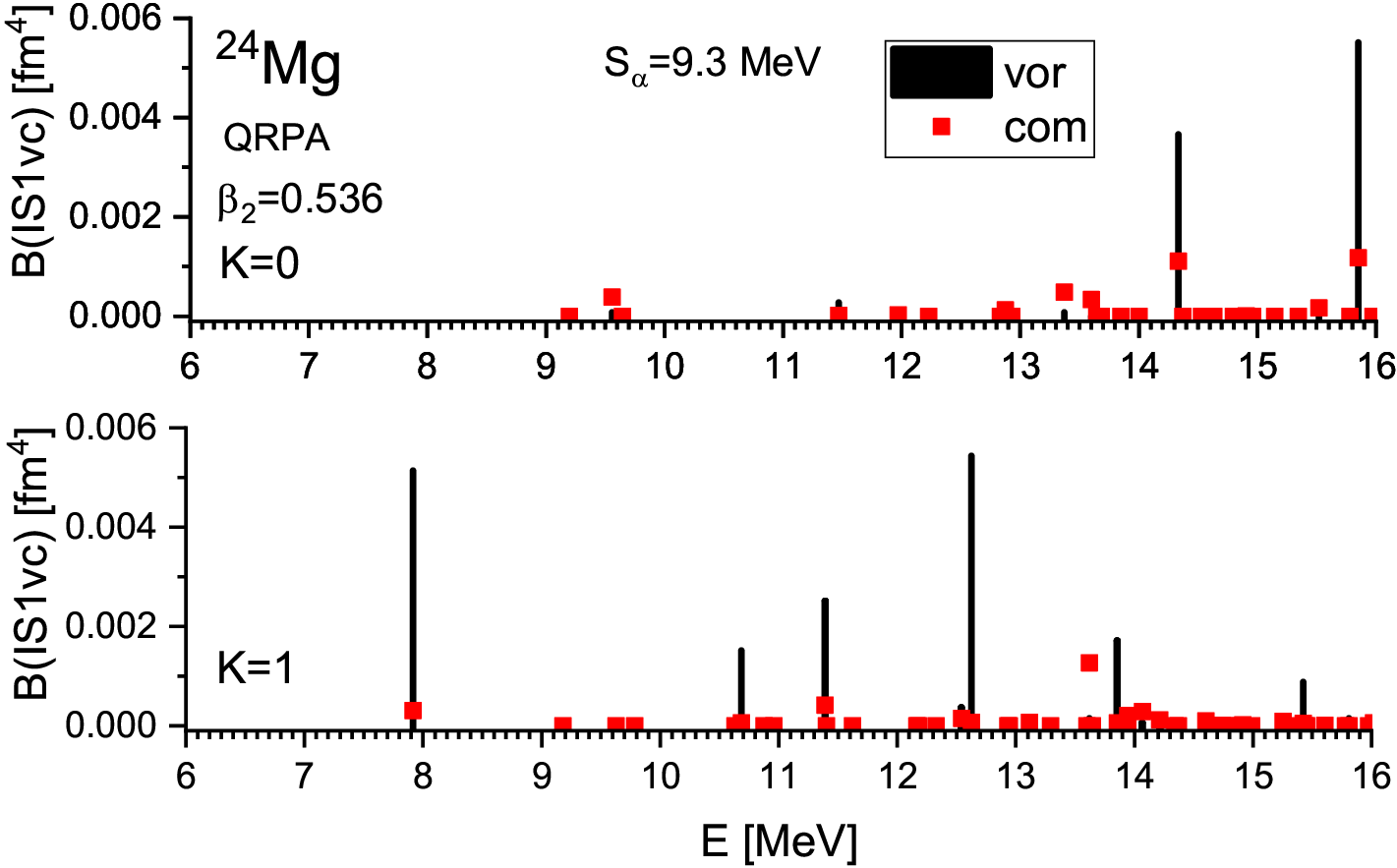}
\caption{QRPA results for isoscalar vortical (black bars) and compression (red
filled squares) dipole strengths in $K=0$ (middle) and $K=1$ (bottom)
dipole states in $^{24}$Mg.}
\end{figure}%

\begin{table} 
\caption{QRPA isoscalar vortical $B(IS1v)$ and compression $B(IS1c)$ strengths
(for $K=0$ and $K=1$) summed over the energy interval $E_x=0-16$ MeV.}
\label{tab-2}
\begin{ruledtabular}
\begin{tabular}{ccc|cc}
 & \multicolumn{2}{c}{$K=0$} & \multicolumn{2}{c}{$K=1$}
\\
\hline
Nucleus  & $B(IS1v)$ & $B(IS1c)$ & $B(IS1v)$ & $B(IS1c)$
\\
\hline
$^{24}$Mg & 0.010  & 0.0038 & 0.019 & 0.0033
\\
$^{26}$Mg & 0.012  & 0.0011 & 0.028 & 0.0074
\\
$^{28}$Si &  0.015  & 0.0011 & 0.029 & 0.0071
\\
\end{tabular}
\end{ruledtabular}
\\
\end{table}

In Figure 6, the vortical $B(IS1v)$ and compression $B(IS1c)$ strengths for $K=0$ and $K=1$ dipole branches in $^{24}$Mg are compared. The states with $B(IS1v)> B(IS1c)$ should be considered as vortical in nature, see e.g. the $E_x = 7.92$-MeV $K=1$ state. Instead, the states with $B(IS1v) < B(IS1c)$ are basically of compressional irrotational character.  Compressional states
can be directly excited in the $(\alpha,\alpha^\prime)$ reaction \cite{harakeh2001giant}. The vortical states usually have a minor irrotational admixture and, most probably, are weakly excited in the $(\alpha,\alpha^\prime)$ reaction through this admixture. Figure 6 shows that, in accordance with previous QRPA predictions \cite{PhysRevLett.120.182501,PhysRevC.100.064302}, the lowest $K=1$ state at $E_x = 7.92$-MeV is mainly vortical. Moreover, for $E_x < 14$ MeV, the $K=1$ branch exhibits much more vorticity than the $K=0$ branch. The summed $B(IS1v)$ and $B(IS1c)$ are reported in Table \ref{tab-2}.

The vortical character of the {\it lowest} dipole state may be a unique peculiarity of $^{24}$Mg. At least, this is not the case in $^{26}$Mg and $^{28}$Si, as discussed below. As mentioned above, the vortical 7.92-MeV $K=1$ state in $^{24}$Mg  is mainly formed by the proton  $pp[211\uparrow - 330\uparrow]$ and neutron $nn[211\uparrow - 330\uparrow]$  $1ph$ configurations.
Just these configurations produce the vortical flow \cite{Ne_EPJ}. The large prolate deformation in $^{24}$Mg downshifts the energy of these configurations, thus making the vortical dipole state the lowest in energy \cite{PhysRevLett.120.182501,Ne_EPJ}. It is remarkable that the previous AMD+GCM calculations \cite{chiba2019cluster} give a very similar result for $^{24}$Mg: that the lowest dipole state at $E_x = 9.2$ MeV has vortical ($K=1$) character and a higher compressional ($K=0$) state at $E_x =  11.1$ MeV.

\subsubsection{$^{26}$Mg}

In Figure 7, we compare the calculated $B(IS1K)$ and $B(IS3K)$ responses with the $(\alpha,\alpha^\prime)$ data. In both experiment and theory, we see numerous dipole states above $E_x \sim 6$ MeV. The fragmentation of the dipole and octupole strengths is somewhat larger than in $^{24}$Mg, which can be explained by the stronger neutron pairing in $^{26}$Mg (in contrast,
the proton pairing in $^{26}$Mg and both proton and neutron pairings in $^{24}$Mg are weak).

Again we see rather large $B(IS3K)$ values, which means that many of the $K=0$ and $K=1$ excitations are of a mixed dipole-octupole character. As in $^{24}$Mg, the states can be separated into two groups, below and above the threshold ($S_{\alpha}=10.6$~MeV). We observe a near-threshold collective $E_x = 9.96$-MeV $K=0$ state with an impressive$B(IS30)$ value.

\begin{figure} 
\includegraphics[width=9cm,height=6cm]{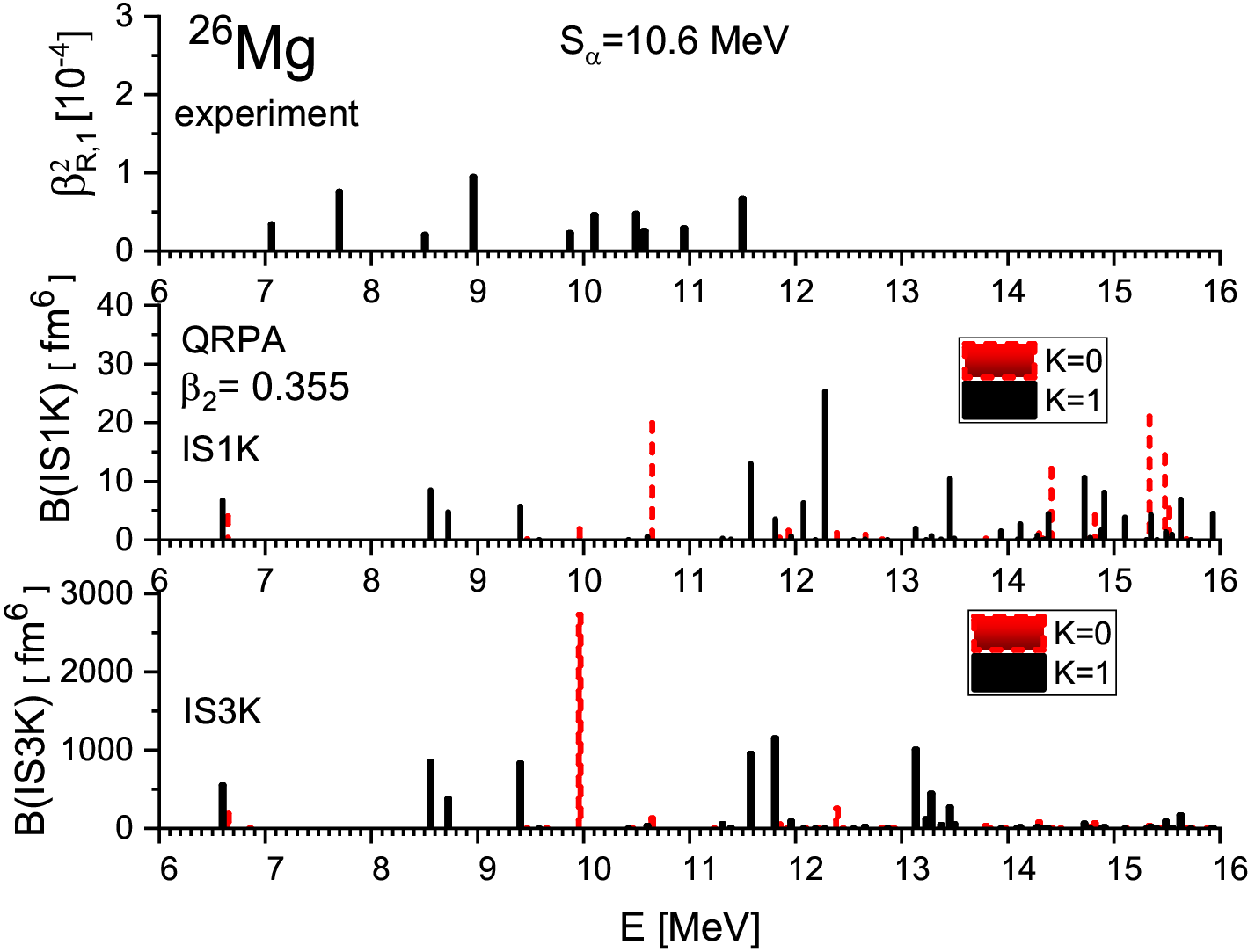}
\caption{The same as in Figure 5 but for $^{26}$Mg.}

\label{fig:26Mg_S_QRPA}
\end{figure}

\begin{figure} 
\includegraphics[width=8.5cm,height=5.5cm,clip]{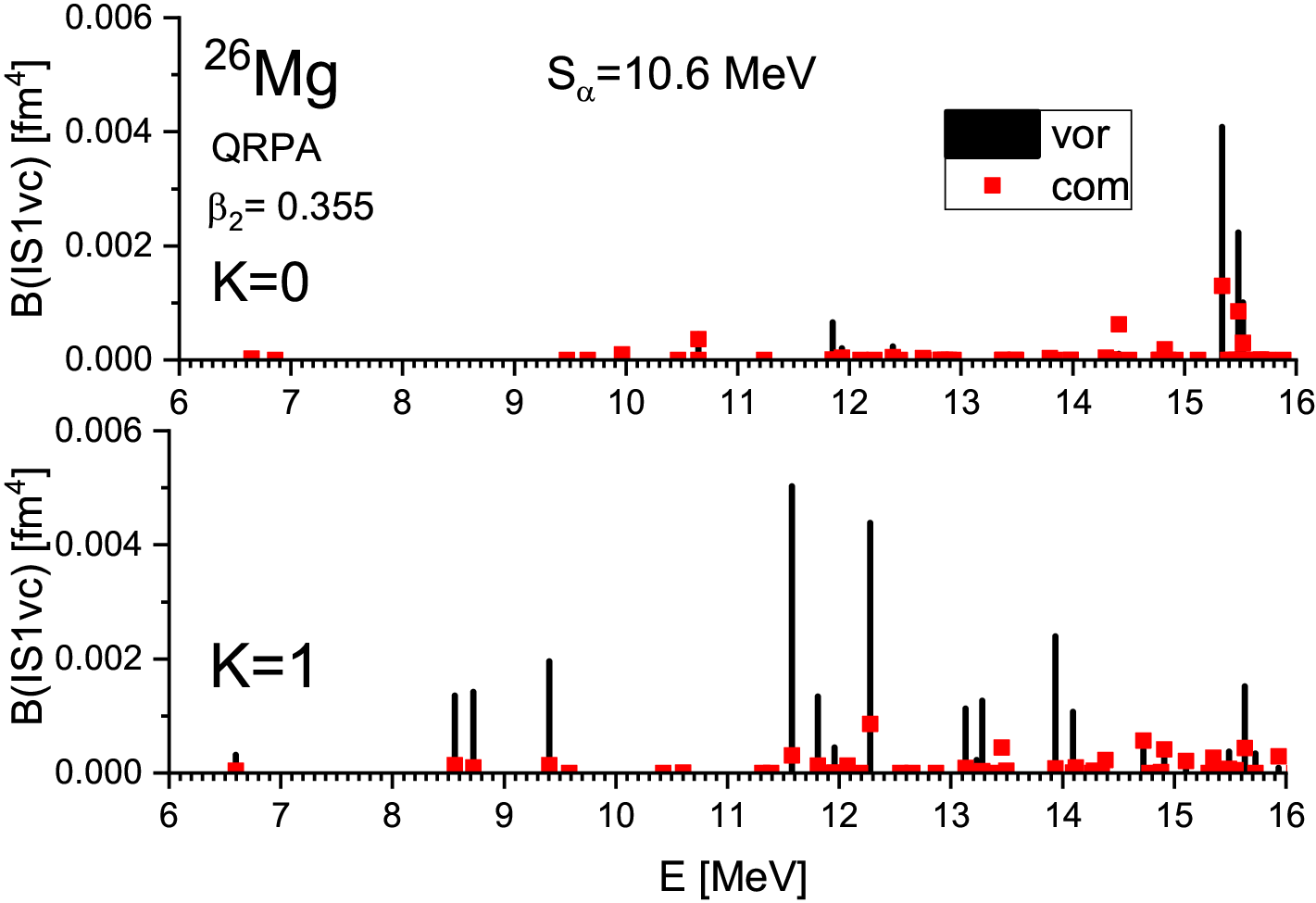}
\caption{The same as in Figure 6 but for $^{26}$Mg.}
\label{fig:26Mg_vc}
\end{figure}

As can be seen in Figure \ref{fig:26Mg_S_QRPA}, the theory suggests another pattern for the lowest dipole states in $^{26}$Mg. Unlike $^{24}$Mg, where the lowest dipole $K=1$ state is
well separated and exhibits a vortical character, the QRPA dipole spectrum
in $^{26}$Mg starts with two almost degenerate $K=1$ and $K=0$ states at $E_x\sim6.6$~MeV. Moreover, as can be seen in Figure \ref{fig:26Mg_vc}, these lowest QRPA states
in $^{26}$Mg are not vortical.


To understand these results, we should inspect the structure of the lowest 6.60-MeV K=1 and 6.64-MeV $K=0$ QRPA states in $^{26}$Mg. They are dominated by $1ph$ neutron configurations $nn[211\downarrow +330]\uparrow$ and $nn[211\downarrow - 330]\uparrow$, respectively. The same content explains the quasi-degeneracy of these states. These $1ph$ configurations have low $B(E1v)$ values and are not vortical. The configurations correspond to $F+1 \to F+5$ transitions, where $F$ marks the Fermi level. Both single-particle levels involved in the transition lie above the Fermi level and the transition is active only because of the developed neutron pairing in $^{26}$Mg (but it is suppressed in $^{24}$Mg, where the calculated pairing is negligible).

Note that $1ph$ excitations $pp[211\uparrow - 330]\uparrow$ and $nn[211\uparrow - 330]\uparrow$, which produce the vorticity in the lowest $K=1$ vortical dipole state in $^{24}$Mg, also exist in $^{26}$Mg, but they are located at a higher energy of $E_x = 8.5-9.5$ MeV. Therefore, the distribution of the vorticity is mainly determined by the energy of vortical $1ph$ configurations.  Besides, it is affected by pairing factors and residual interaction.

\subsubsection{$^{28}$Si}

\begin{figure} 
\includegraphics[width=8.5cm,height=6.5cm,clip]{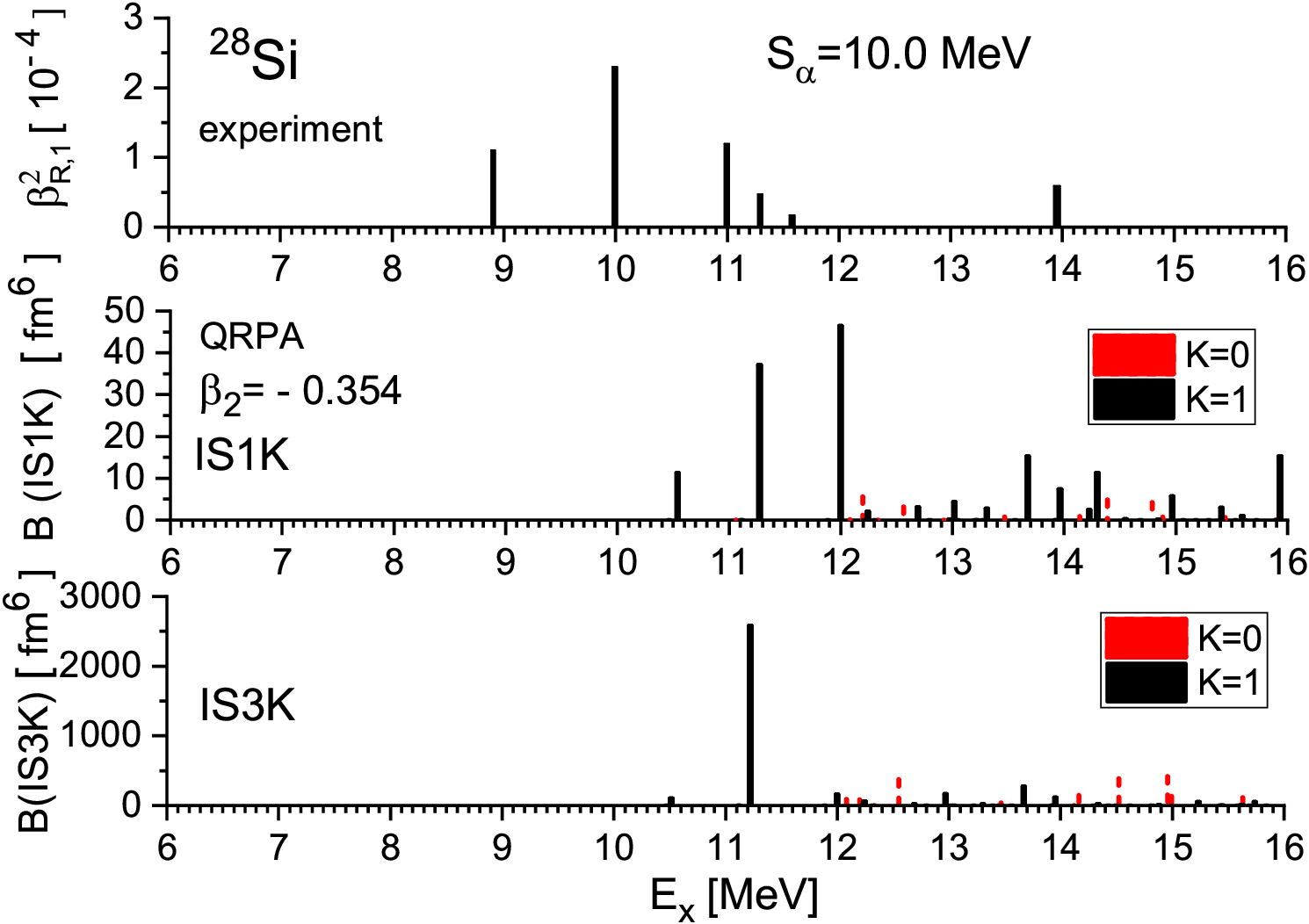}
\caption{The same as in Figure 5 but for $^{28}$Si.}
\label{fig:28Si_S_QRPA_AMD}
\end{figure}

In Figure \ref{fig:28Si_S_QRPA_AMD}, we present the experimental data and QRPA results for $IS1K$ and $IS3K$  strengths in oblate $^{28}$Si. We see that the theory
significantly overestimates the energy of the lowest $K^-$ state: it appears at 8.8 MeV in experiment and at 10.5 MeV in QRPA. So, unlike the experiment, the theory does not suggest any $K^-$ states  below the threshold ($S_{\alpha}=9.98$~MeV). Perhaps this discrepancy is caused by a suboptimal oblate deformation $\beta_2= -0.354$ used in our calculations. Further, Figure \ref{fig:28Si_S_QRPA_AMD} and Table \ref{tab-1} show that the dipole and octupole strengths for $K=1$ are much larger than for $K=0$. So, in this nucleus $K=1$ states should be more strongly populated in $(\alpha,\alpha^\prime)$ than $K=0$ states.

The bottom panel of Figure \ref{fig:28Si_S_QRPA_AMD} shows that, with the exception of the $E_x = 11.2$-MeV $K=1$ state, the nucleus $^{28}$Si does not demonstrate any fundamental octupole strength. So, for most of its $K^-$ states, the dipole-octupole coupling is suppressed.  The near-threshold state at 11.2 MeV with significant octupole strength has $K=1$ but not $K=0$ as in $^{24,26}$Mg. Perhaps all these peculiarities are caused by the oblate deformation of $^{28}$Si.

\begin{figure} 
\includegraphics[width=8.5cm,height=5.5cm,clip]{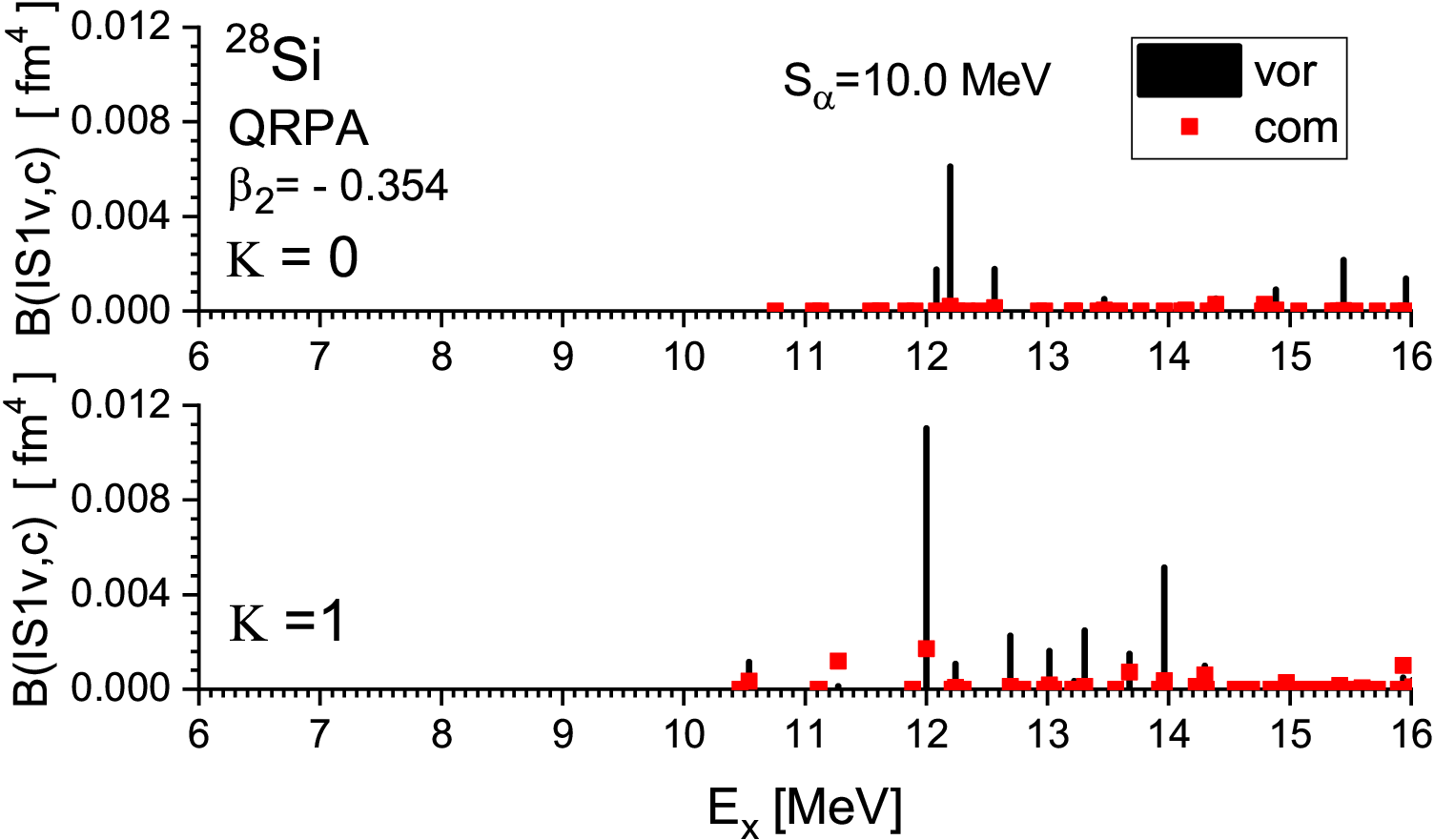}
\caption{The same as in Figure 6 but for $^{28}$Si.}
\label{fig:28Si_vc}
\end{figure}

In our calculations, the pairing in $^{28}$Si is weak. As a result, the vortical configuration  $[211]\uparrow - [330]\uparrow$ corresponding in this nucleus to the transition between particle states is suppressed. So, as seen from Figure \ref{fig:28Si_vc}, the lowest dipole states in $^{28}$Si are not vortical and vorticity appears only above 12 MeV. As in $^{24,26}$Mg, the vorticity is mainly concentrated in the$K=1$ branch.


\begin{figure*}
    \centering
    \includegraphics[width=0.7\textwidth]{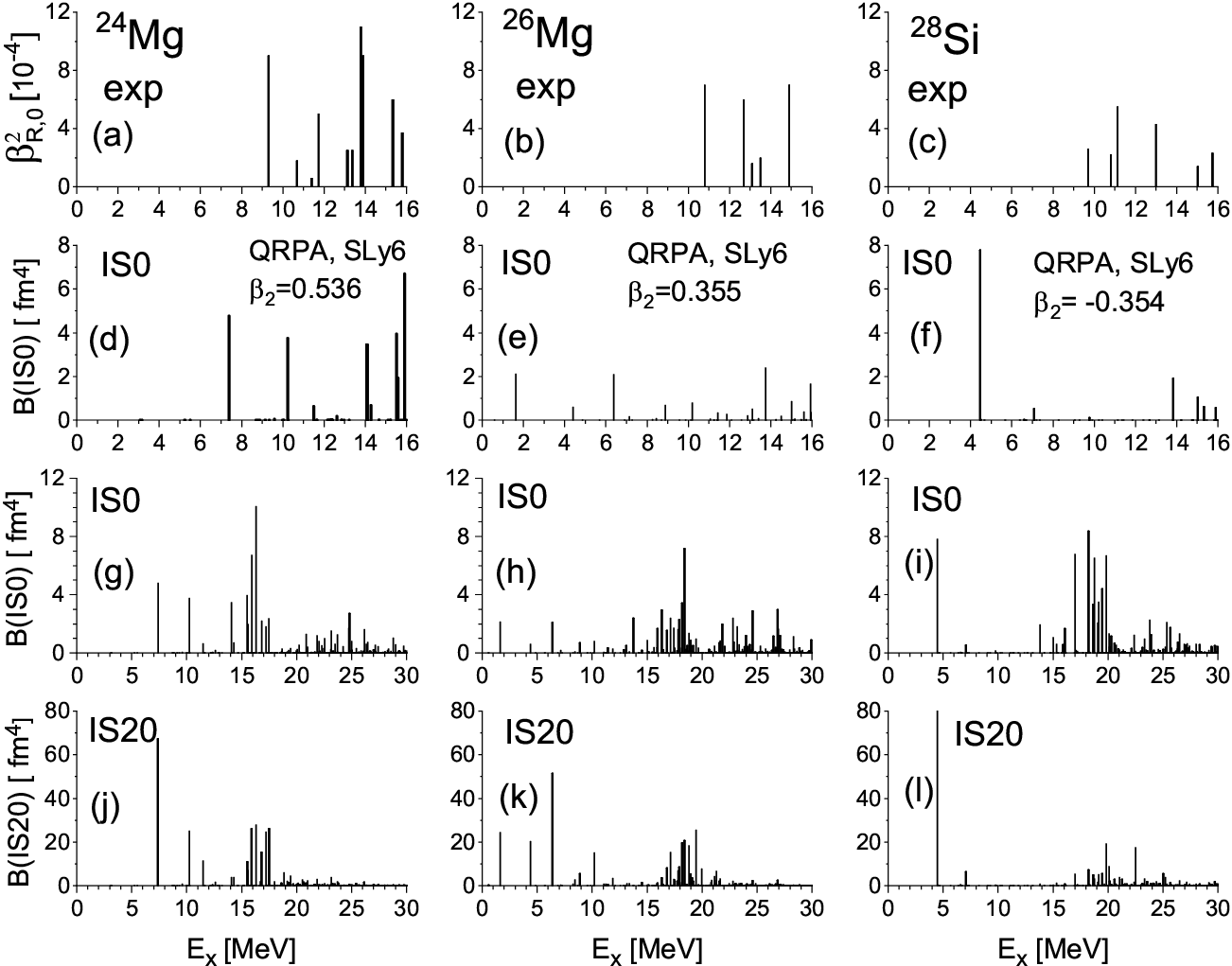}
    \caption{Experimental $\beta^2_{R,0}$ factors (a-c), QRPA $B(IS0)$ values
    for $K^{\pi}=0^+$ excitations at 0-16 MeV (d-f) and 0-30 MeV  (g-i), QRPA
    $B(IS20)$ values at 0-30 MeV (j-l).}
    \label{fig:E0_all}
\end{figure*}

\subsection{Summary for $IS1$ QRPA results}

QRPA calculations do not allow one to establish a direct correspondence between the calculated and observed $1^-$ states. Perhaps this is because the present QRPA scheme does not take into account such important factors as triaxiality, shape coexistence, clustering and complex configurations are omitted. Nevertheless, the QRPA calculations lead to some interesting and robust results.

\begin{enumerate}
\item The strong deformation-induced mixture of the dipole and octupole modes is predicted  for most of  $K^{\pi}=0^-$ and $1^-$ states in $^{24,26}$Mg and in a few particular states in $^{28}$Si. Some mixed states demonstrate impressive octupole transition probabilities $B(IS3K)$. Perhaps these states belong to the low-energy octupole resonance (LEOR)\cite{harakeh2001giant,Malov76}.

\item In all three nuclei, the collective state with a large octupole strength is predicted near the $\alpha$-particle thresholds $S_{\alpha}=9.3-10.6$ MeV. This state has $K=0$ in $^{24,26}$Mg and $K=1$ in $^{28}$Si. Most probably, the difference is caused by different signs of the axial deformation in these nuclei.

\item Above the $\alpha$-particle thresholds, fragmented vorticity is found in $K=1$ states in all three nuclei. Below $S_{\alpha}$, the picture is different: the vorticity is concentrated in the lowest dipole state at $E_x\sim 8$ MeV in $^{24}$Mg, fragmented between several states at $E_x\sim 8.5-9.5$ MeV in $^{26}$Mg, and fully absent in $^{28}$Si. As was discussed,
the vorticity is delivered by particular $1ph$ configurations which can have a different energy location depending on the nuclear deformation and other factors, e.g. the residual interaction. Moreover, these configurations are active only if they are of particle-hole character or supported by the pairing (like in $^{26}$Mg). A particular interplay of these  factors in $^{24,26}$Mg and $^{28}$Si leads to the difference in their vorticity distribution.

\end{enumerate}

\subsection{$IS0$ strength distributions}

In Figure \ref{fig:E0_all}, the  $(\alpha,\alpha^\prime)$ data for $K^{\pi}=0^+$ states in $^{24,26}$Mg  and $^{28}$Si (plots (a)-(c)) are compared with QRPA isoscalar monopole strengths ($B(IS0)$) in the energy interval 0-16 MeV (plots (d)-(f)).

As mentioned above, because of the limitations of the experimental set up,
the present $(\alpha,\alpha^\prime)$ data cover $E_x = 9-16$ MeV. Low-energy $0^+$ states listed in Tables I, III and V of Section IV are omitted in Figure \ref{fig:E0_all}.

Figure \ref{fig:E0_all} shows that in $^{24}$Mg the experimental and QRPA strength distributions look rather similar. The calcu-
lated $E_x = 7.38$-MeV state perhaps corresponds to the observed $E_x = 6.41$-MeV state \cite{PhysRevC.93.044324,GUPTA2015343}.
In $^{26}$Mg, the situation is quite different since QRPA predicts $IS0$ states from around $E_x = 1-2$ MeV. In $^{28}$Si, QRPA suggests the onset of $0^+$ states around $E_x = 4-6$ MeV. In all three nuclei, QRPA predicts some $0^+$ states at 9-16 MeV, which is in general accord with the experimental $(\alpha,\alpha^\prime)$ data. The QRPA IS0 strengths summed over $E_x = 0-16$ MeV are 26.1 fm$^4$, 13.76 fm$^4$, and 12.6 fm$^4$ in $^{24}$Mg, $^{26}$Mg, and $^{28}$Si, respectively.

Note that, in the QRPA calculations,  the actual number of $0^+$ states at $E_x < 16$ MeV is much larger than might be seen in Figure \ref{fig:E0_all}. In fact, QRPA gives
48 ($^{24}$Mg), 53 ($^{26}$Mg), and 50 ($^{28}$Si) states. However, most of these states are not seen in plots (d)-(f) because of their very small $B(IS0)$ values.

Plots (g)-(i) in Figure \ref{fig:E0_all} show QRPA $B(IS0)$ strength in the larger energy interval 0-30 MeV including the IsoScalar Giant Monopole Resonance (ISGMR). In deformed nuclei, there is the  coupling of monopole and quadrupole modes, see e.g. early studies \cite{harakeh2001giant,Abg80,Jang83} and recent systematic studies \cite{Garg_rew19,Nest_PRC16}.  In particular, the ISGMR is coupled with the $\lambda\mu=20$ branch of the IsoScalar Giant Quadrupole Resonance, ISGQR($20$). Due to this coupling, a part of the $IS0$ strength is transferred from the energy region of the normal ISGMR to the energy region where the ISGQR($20$) branch is located. Thus, we get the deformation-induced splitting of the ISGMR strength into two parts, the main ISGMR
fraction, and additional strength located at the energy of the ISGQR($20$) component. Since the ISGQR lies below the ISGMR, this strength also appears below the ISGMR. The larger the deformation, the more $IS0$ strength is transferred to this lower fragment from the main ISGMR, see Ref. \cite{Nest_PRC16} for more detail. For light nuclei of our present interest, the deformation-induced coupling of monopole and quadrupole modes was earlier studied using ($\alpha,\alpha^\prime$) reaction for $^{24}$Mg \cite{PhysRevC.93.044324,GUPTA2015343} and $^{28}$Si 
\cite{PhysRevC.93.064325}.

In our calculations, $^{24,26}$Mg and $^{28}$Si have large quadrupole deformations and so we should expect significant ISGMR splitting. Indeed, the plots (g)-(i) show that the  ISGMR in these nuclei is split into two main sections: the narrow distribution between 15 and 19 MeV and the main, wider, ISGMR distribution between 20 and 30 MeV. The picture is similar in prolate $^{24,26}$Mg and oblate $^{28}$Si. Note that the obtained distributions of IS0 strength rather well reproduce the experimental data for $^24$Mg \cite{PhysRevC.93.044324,GUPTA2015343} and $^{28}$Si\cite{PhysRevC.93.064325}.

The above  treatment of ISGMR splitting is justified by the plots (j)-(l), where the strength $B(IS20)$ of quadrupole isoscalar transitions $0^+0_{gs} \to 2^+0_{\nu}$ from the ground state to the rotational quadrupole state built on the band-head $|\nu\rangle$ is exhibited. We see that the ISGQR($20$) branch is located at 15-19 MeV, i.e. at the same energy as the narrow $IS0$ hump.
This confirms that the $IS0$ hump is just the ISGMR part arising due to the deformation-induced ISGMR/ISGQR coupling realized for  $K^{\pi}=0^+$ states.

Plots (g)-(i) and (j)-(l) highlight some important points. First, the plots (g)-(i) show that the $J^\pi = 0^+$ states in $(\alpha,\alpha')$ data lie just below the ISGMR peak, i.e. basically beyond the ISGMR. Only in  $^{24}$Mg, these states perhaps cover the edge of the ISGMR hump. Second, from comparison of the plots (g)-(i) and (j)-(l), we learn that $K^{\pi}=0^+$ states at 0-16 MeV exhibit both strong $IS0$ and $IS20$ transitions. They should, therefore, not be treated as solely monopole states but rather as strong mixtures of monopole and quadrupole excitations.

\section{Comparison with AMD+GCM calculations}
\label{sec:Comparison}

In this section, we discuss the comparison between the present experimental results and the AMD+GCM calculations for $^{24}{\rm Mg}$ and $^{28}{\rm Si}$ presented in
Refs.~\cite{Kimura_E0_24Mg,PhysRevC.95.044328,Taniguchi2020}. These calculations do not take into account all the degrees of freedom of the collective excitations. Therefore, they are not appropriate for the discussion of the global features of the observed strength distributions. However, AMD+GCM describes the clustering aspects which involves many-particle-many-hole excitations, and hence, can offer a different insight into the low-lying strengths than that from QPRA. From the mean-field side, AMD+GCM takes into account the interplay between axial and triaxial nuclear shapes, which is important for light nuclei.

In Ref. \cite{Kimura_E0_24Mg}, using the AMD+GCM framework, the relationship between the monopole strengths in $^{24}{\rm Mg}$ and clustering has been discussed. The $\alpha$+$^{20}{\rm Ne}$, $^{8}$Be$+^{16}$O, $^{12}$C$+^{12}$C and $5\alpha$ cluster configurations were investigated in addition to the $1ph$ single-particle excitations. It was concluded that several low-lying monopole transitions at energies below the giant monopole resonance can be  attributed to the clustering as summarized in Table \ref{tab:amd1}.

\begin{table*}[hbt!]
  \caption{The cluster configurations with significant $B(IS0)$ strengths and their excitation energies $E_x$ in $^{24}{\rm Mg}$ calculated by AMD+GCM and compared with the observed data
  \cite{ENSDF24Mg}. The energies $E_{\rm shift}$ are obtained by a downshift of 2.9 MeV so as to adjust $E_{\rm exp}$ for the $0_2^+$ state.}
  \label{tab:amd1}
\begin{ruledtabular}
\begin{tabular}{ccccccc}
 Cluster                       & $J^\pi$ & $E_x$ [MeV]  & $B(IS0)$ [fm$^2$] & $E_{\rm shift}$ [MeV] &
 $E_{\rm exp}$ [MeV] & $B(IS0)_{\rm exp}$ [fm$^2$] \\\hline
                               & $0_2^+$ & 9.3    & 9.7    & 6.4             &
 6.4           & 14.3$\pm$ 1.6         \\
 ($^{20}{\rm Ne}$+$\alpha$)      & $0_3^+$ & 11.7   & 4.7    & 8.8             &
 9.3           &                   \\
 $^{20}{\rm Ne}$+$\alpha$      & $0_5^+$ & 13.2   & 2.5     & 10.3            &
               &                   \\
 $^{12}{\rm C}$+$^{12}{\rm C}$ & $0_8^+$ & 15.3   & 6.2    & 12.4            &
               &
\end{tabular}
\end{ruledtabular}
 \end{table*}

As already discussed in previous works on AMD+GCM and QPRA calculations \cite{Taniguchi2020,Peru2008,Peru2014}, the Gogny D1S interaction overestimates the energy of the non-yrast states of $^{24}{\rm Mg}$. Therefore, when we compare the AMD+GCM results listed in Table \ref{tab:amd1} with the experiment, it is better to shift down the calculated excitation energies to match with the well-known states. For this purpose, Table \ref{tab:amd1} also lists the calculated excitation energies shifted down by 2.9 MeV so as to reproduce the observed energy ($E_{\rm exp}$=6.4 MeV) of the $0_2^+$ state. Note that this shift also changes the calculated excitation energy of the $0_3^+$ state (11.7 MeV$\rightarrow$ 8.8 MeV) close to the observed value of $E_x = 9.3$ MeV which is experimentally well established. For the higher excited states (the $0_5^+$ and $0_8^+$ states), as the observed level density is rather high, the experimental counterparts in the ENSDF database \cite{ENSDF24Mg} are ambiguous.

Table \ref{tab:amd1} should be compared with the present experimental data from Table \ref{tab:Mg24_J_eq_0_states} and Figure \ref{fig:E0_all}. We see that the $0_2^+$ state is out of the acceptance of the present experiment, but the $0_3^+$ state is clearly observed and has the enhanced monopole strengths as predicted by AMD+GCM. In Ref.~\cite{Kimura_E0_24Mg}, it was concluded that $0^+_3$ is a mixture of the collective and $^{20}{\rm Ne}+\alpha$ cluster excitations. Consequently, it is  interesting to note that the $0^+_3$ also appears as a prominent peak in the QRPA result (Figure \ref{fig:E0_all}). In addition to the $0^+_3$ state, Table \ref{tab:Mg24_J_eq_0_states} reports a state at 11.7 MeV and a group of states at 13.0-13.9 MeV with the enhanced monopole strengths. These states are of particular interest because their energies are close to the corresponding cluster decay thresholds (9.3 MeV for $^{20}{\rm Ne}$+$\alpha$, 13.9 MeV for $^{12}{\rm C}$+ $^{12}{\rm C}$, 14.047 MeV for $^{16}$O$+2\alpha$ and 14.138 MeV for $^{16}$O$+^8$Be as listed in the Ikeda diagram \cite{Ikeda1968}. Furthermore, these states are also visible in the excitation function reported in another $^{24}$Mg($\alpha,\alpha^\prime$)$^{24}$Mg experiment and seem not be reproduced by RPA calculations \cite{PhysRevC.93.044324, Peru2008}. Therefore, they can be attributed to the cluster resonances. In the AMD+GCM calculations, the candidates of the $^{20}{\rm Ne}$+$\alpha$ and $^{12}{\rm C}$+ $^{12}{\rm C}$ cluster configurations were predicted at 13.2 and 15.3 MeV (10.3 and 12.4 MeV with the 2.9-MeV shift),
respectively. Of course, to firmly establish the assignments of these states, more detailed analysis is indispensable. For example, the differential cross sections of these states should be compared with theoretical predictions in the future. The present experiment probes only a small range of angles and is insufficient for thorough comparison with theory.

For $^{28}{\rm Si}$, AMD+GCM calculations suggest pairs of $0^+$ and $1^-$ states pertinent to asymmetric cluster configurations, such as $^{24}{\rm Mg}$+$\alpha$, $^{20}{\rm Ne}$+$^{8}{\rm Be}$ and $^{16}{\rm O}$+$^{12}$ \cite{PhysRevC.95.044328}. The predicted results are summarized in Table \ref{tab:amd2}. Similar to the $^{24}{\rm Mg}$ case, the Gogny D1S interaction systematically overestimates the energies of the non-yrast states, see Figure 6 in Ref.~\cite{PhysRevC.95.044328}. Therefore, while comparing the AMD+GCM and experimental results, we again use the downshift of the calculated excitation energies, now  by 3.3 MeV, to match the energy of the observed $0_3^+$ state. Note that this well-known prolate-deformed state should have a
large contribution from the $^{16}{\rm O}$+$^{12}{\rm C}$ cluster configuration \cite{Baye1976,Baye1977,Taniguchi2009,PhysRevC.86.064309}. The value of the energy
downshift looks reasonable as it is similar to that introduced for $^{24}{\rm Mg}$. With this shift, the energies of other well-known states show the reasonable agreement between the AMD+GCM and experimental results. For example, the $2^+$ member of the SuperDeformed (SD) band, which has been experimentally identified at 9.8 MeV in Ref.~\cite{PhysRevC.86.064308}, agrees well with the shifted AMD+GCM state at 9.7 MeV. Furthermore, a couple of the $^{24}{\rm Mg}$+$\alpha$ cluster resonances have been identified around 13 MeV in resonant scattering
experiments \cite{Tanabe1983,Artemov1990}, which are close to the shifted AMD $0_6^+$ state at 14.9 MeV.

\begin{table}
\caption{The cluster configurations with their excitation energies
$E_x$ and transition strengths ($B(IS0)$ for the $0^+$ states and
$B(IS1)$ for the $1^-$ states) in $^{28}{\rm Si}$, calculated within
AMD+GCM. The experimental counterparts are taken from Ref.
 \cite{Basunia2013} and the present experiment (denoted
 by bold).  The energies $E_{\rm shift}$ are obtained by a downshift of
 3.3 MeV so as to adjust the energy $E_{\rm exp}$=6.69 MeV for the $0_3^+$ state.}
 \label{tab:amd2}
\begin{ruledtabular}
\begin{tabular}{ccccccc}
 cluster                        & $J^\pi$        & $E_x$ & $B(IS\lambda)$ & $E_{\rm shift}$
 & $E_{\rm exp}$ & $B(IS0)_{\rm exp}$\\\hline
                                & $0_2^+$        & 5.8   & 16.0           & 2.5
 & 4.98          & 14.7\\
 $^{20}{\rm Ne}$+$^{8}{\rm Be}$ & $0_5^+$        & 13.8  & 9.3           & 10.5\\
                                & $1_2^-$ & 14.9  & 90.3           & 11.6\\
 $^{24}{\rm Mg}$+$\alpha$       & $1_1^-$  & 12.9  & 130.0          & 9.6 \\
                                & $0_6^+$        & 18.2  & 5.1           & 14.9
 &{\bf 13.0}\\
                                & $1_5^-$  & 20.6  & 64.0           & 17.3\\
                                &                & 21.5  & 1.7           & 18.2\\
                                &                & 22.5  & 6.8           & 19.2\\
 $^{12}{\rm C}$+$^{12}{\rm C}$  & $0_3^+$        & 10.0  & 0.0           & 6.7
 &6.69\\
		                & $1_3^-$& 15.8  & 0.0           & 12.5\\
 $^{24}{\rm M}$+$\alpha$ (SD)   & $0_4^+$        & 12.6  & 0.0           & 9.3
 &{\bf 9.7}\\
                                & $2_5^+$        & 13.0  &               & 9.7
 &9.8\\
		                & $1_4^-$ & 17.6  & 0.0           & 14.3\\
                                &                & 18.8  & 0.0           & 15.5
\end{tabular}
\end{ruledtabular}
\end{table}

We now examine the cluster configurations listed in Table \ref{tab:amd2} and compare to the present experimental data. Since the monopole ($IS0$) and dipole $(IS1)$ transitions have a
strong selectivity for the cluster states, the cluster configurations can be classified into two groups which are strongly populated/hindered in the $(\alpha,\alpha^\prime)$ reaction. For example, from a simple theoretical consideration, we can predict that the $0_3^+$ state that is the band head of the prolate band (the lowest $^{16}{\rm
O}$+$^{12}{\rm C}$ cluster  band) should be hindered. See Ref.~\cite{1742-6596-863-1-012024} for details of the hindrance  mechanism. It  is interesting that the hindrance of the $0^+_3$ state can also be seen in the QRPA results shown in Figure \ref{fig:E0_all}. Unfortunately, this state (which is important for validation of the relationship between the monopole transitions and clustering) is out of the acceptance of the present experiment but it should be experimentally confirmed to validate the discussion the hindrance of the transition.

For the same reason, the AMD+GCM predicts that the SD band head expected at 9.3 MeV should also be hindered. However, in the present experiment, the observed 9.7-MeV $0^+$ state is very close to the 9.8-MeV $2^+$ state and, following Table V, has the enhanced monopole strength in contradiction to the AMD+GCM prediction. This new result requires a more detailed analysis of the SD state in $^{28}$Si.

At the same time, AMD+GCM predicts an enhancement of the $^{20}$Ne$+^{8}$Be and $^{24}$Mg$+\alpha$ cluster configurations. The pair of the $0_5^+$ and $1_2^-$ states with the $^{20}$Ne$+^{8}$Be configuration is predicted at$E_x = 10-11$ MeV, and some fractions of $IS0$ and $IS1$ strength are indeed experimentally observed in this energy region.
This may be the first indication of the $^{20}$Ne$+^{8}$Be clustering in $^{28}{\rm Si}$, which must be confirmed by a more detailed study, e.g. the transfer of $^{8}{\rm Be}$ to $^{20}{\rm Ne}$. Other states which are predicted to be strongly populated in the $(\alpha,\alpha^\prime)$ reaction are $^{24}$Mg$+\alpha$ cluster states. AMD+GCM calculations predict a $1_1^-$ state at $E_x = 9.6$ MeV and $0^+$ and $1^-$ states at approximately $15$ and $17-20$ MeV.
The $1^-$ states at 17-20 MeV are beyond the present experiment. Several $0^+$ states can be seen at 9.5 MeV and 15 MeV. It is worthwhile to note that the $\alpha$ transfer and $\alpha$+$^{24}{\rm Mg}$ resonant scattering experiments \cite{Tanabe1983,Artemov1990} also report a group of the $\alpha$+$^{24}{\rm Mg}$ resonances with $J=0^+$ within  the same energy region. Therefore, the data of the previous and present experiments  as well as the AMD+GCM results look consistent. A more detailed comparison between AMD+GCM and experimental results may be conducted in the future.

\section{Conclusions}
\label{sec:Conclusions}

The  isoscalar dipole ($IS1$) and monopole ($IS0$) excitations of $^{24}$Mg, $^{26}$Mg and $^{28}$Si at the energy interval $E_x = 9-16$ MeV have been measured using the ($\alpha,\alpha^\prime$) inelastic-scattering reaction at forward angles (including zero degrees). The experiment was performed using the K600 magnetic spectrometer at iThemba LABS (Cape Town, South Africa). New monopole and dipole states were reported.

The extracted $IS1$ and $IS0$ strength distributions were compared to the theoretical calculations performed within the Skyrme Quasiparticle Random-Phase-Approximation (QRPA) \cite{BenRein2003,repko2015skyrme,Repko2017,PhysRevC.99.044307} and Antisymmetrized Molecular Dynamics + Generator Coordinate Method (AMD+GCM) \cite{PhysRevC.95.064319,10.1093/ptep/ptz049,PhysRevC.97.014303,PhysRevC.100.014301,chiba2019cluster} approaches. The correspondence, at least tentative, between some calculated and observed states was established. This theoretical analysis allows us to draw some important physical conclusions.

First of all, QRPA and AMD+GCM calculations  suggest that low-lying $IS1$ states in light nuclei can have two origins: irrotational cluster (IC) \cite{PhysRevC.95.044328}
and mean-field (MF) \cite{PhysRevLett.120.182501,Ne_EPJ,chiba2019cluster}. The MF-states can be irrotational (IMF) and vortical (VMF) \cite{PhysRevLett.120.182501,Ne_EPJ,chiba2019cluster}.

The IC states produce $T=0$ negative-parity cluster bands, which are the doublets of the positive-parity bands based on the monopole states \cite{PhysRevC.95.044328}.
Some traces of these doublets were found in the comparison of theoretical calculations and experimental data. IC states are irrotational dipole oscillations of the two clusters which constitute the nucleus relative to one other. These states originate from the reflection-asymmetric form of the nucleus exhibiting the clustering. The negative-parity bands produced by IC states usually have $K=0$. 

Instead, the VMF states in light nuclei were predicted in the papers of Nesterenko and Kanada-En’yo. They are vortical (not irrotational) toroidal states and are mainly of mean-field origin \cite{PhysRevLett.120.182501,Ne_EPJ,chiba2019cluster}. In general,they can take place in both light and heavy nuclei and can exist without clustering. They do not need the reflection-asymmetric nuclear shape and the associated the monopole doublets. Following previous studies \cite{PhysRevLett.120.182501,Ne_EPJ,chiba2019cluster} and present QRPA calculations, these states produce negative-parity rotational bands, mainly with $K=1$. 

Both IC and IMF/VMF states exhibit enhanced $IS1$ transitions and are usually located near the alpha-particle threshold. In general, IC and IMF/VMF states can be mixed, especially in soft and triaxial nuclei exhibiting $K$-mixing. Nevertheless, the relation to the $K=0$ or $K=1$ band is perhaps a reasonable indicator for an initial discrimination of IC and VMF states.

Being strongly deformed, $^{24,26}$Mg and $^{28}$Si should exhibit a strong coupling between dipole and octupole modes and between monopole and quadrupole modes. This coupling was confirmed by QRPA calculations where strong $IS3K$ ($0^+0_{gs} \to 3^-K_{\nu}$) and $IS20$  ($0^+0_{gs} \to 2^+0_{\nu}$) transitions were found. So, theoretically explored states are actually dipole/octupole and monopole/quadrupole mixtures. Further, QRPA predicts that, near the $\alpha$-particle threshold, there should exist a specific collective state ($K=0$ in prolate and $K=1$ in oblate nuclei) with an impressive octupole strength. This near-threshold state manifests the onset of states with cluster features.

Due to triaxiality and significant shape coexistence  in $^{24,26}$Mg and $^{28}$Si, QRPA results obtained at the fixed axial deformation should be considered as approximate. In addition, QRPA calculations do not include all the dynamical correlations coupling with complex configurations. Nevertheless, the main QRPA prediction - of vortical dipole states with enhanced $IS1$ strength as an alternative to the cluster dipole states - remains robust. In our opinion, more involved calculations may change some details but not this general prediction.

Another interesting QRPA prediction is a change in dipole vorticity below the $\alpha$-particle thresholds in $^{24,26}$Mg and $^{28}$Si. Following our analysis, the vorticity is concentrated in the lowest dipole state in $^{24}$Mg at $\sim$ 8 MeV, is fragmented between several states at $\sim$ 8.5-9.5 MeV in $^{26}$Mg, and is fully absent in $^{28}$Si. The difference is explained by the different energies of $1ph$  configurations responsible for the vorticity. Our explorations confirm the suggestion made in Ref. \cite{PhysRevLett.120.182501} that $^{24}$Mg is perhaps the unique nucleus with a well-separated low-energy vortical state.

In some particular cases, the correspondence between the observed and calculated low-lying states was established. However neither QRPA nor AMD+GCM are still able to provide a systematic one-to-one correspondence of low-lying spectra and experimental data. This demanding task calls for more involved theories, e.g. taking into account the coupling with complex configurations. 

The present ($\alpha,\alpha^\prime$) data do not yet allow confident assignment of the vortical or cluster character of the excitations. However, these data improve our knowledge of the isoscalar  monopole and dipole states at the excitation energies where the clustering and vorticity are predicted. This is a necessary and important step in the right direction. The use of the ($\alpha,\alpha^\prime$) reaction at intermediate energies complements other suggested mechanisms for populating cluster and vortical states such as the ($\gamma,\gamma^\prime$) \cite{Iachello19851,PhysRevLett.114.192504}, ($e,e^\prime$) \cite{PhysRevC.100.064302} and ($d,^6$Li) reactions \cite{PhysRevLett.114.192504} or ($^{6/7}$Li,$d/t$), although detailed information on the interior of nuclei and the vortical mode is likely only available from the $(e,e^\prime)$ reaction. It was recently shown that vortical states in $^{24}$Mg are characterized by the strong interference between the orbit and spin contributions to the experimentally accessible $(e,e^\prime)$ transversal form factors \cite{PhysRevC.100.064302}. This results in specific momentum distributions for $E1$ (and $M2$ in deformed nuclei) backward scattering, which in turn allows identification of vortical states \cite{PhysRevC.100.064302}. Branching ratios and transition strengths of $\gamma$-ray transitions from the observed dipole states would provide information on the $K$ assignment of the levels and should also be a focus of additional future experimental work.

Modern theoretical methods still cannot provide a comprehensive  description of all the important aspects of light nuclei (clustering, softness, shape coexistence, mean-field features like vortricity, etc.) with an acceptable computational effort.  Thus a comparative analysis with different theoretical methods, e.g. AMD + GCM and QRPA, is presently the best way to proceed.  Additional methods taking into account the coupling with complex configurations, e.g. the shell-model approach, are also welcome.  Between various models, AMD + GCM looks to be the most powerful and promising tool. Indeed, using a sufficiently  large set of basis functions, this model can potentially describe both cluster and mean-field degrees of freedom and take into account the shape coexistence. In addition, the AMD + GCM results are physically transparent. However it is not yet easy to exploit the full potential of AMD + GCM calculations as then we need a large basis set and thus a huge computational effort. At present, the most realistic way is to combine AMD + GCM with other models as was done in our study.
 
 \acknowledgements

The authors thank the Accelerator Group at iThemba LABS for the high-quality dispersion-matched beam provided for this experiment. PA acknowledges support from the Claude Leon Foundation in the form of a postdoctoral fellowship, and thanks M. N. Harakeh for providing the {\sc belgen} and {\sc fermden} codes and helpful advice regarding the DWBA calculations, and Josef Cseh for useful discussions concerning $^{28}$Si . RN acknowledges support from the NRF through Grant No. 85509. VON and JK thank Dr. A. Repko for the QRPA code. The work was partly supported by Votruba - Blokhintsev (Czech Republic - BLTP JINR) grant (VON and JK) and a grant of the Czech Science Agency, Project No. 19-14048S (JK). VON and PGR appreciate the Heisenberg-Landau grant (Germany DLTP JINR).

\appendix

\section{Details of DWBA calculations}
\label{app:DWBA}

In past studies, e.g. \cite{GUPTA2015343,PhysRevC.93.044324,PhysRevC.60.014304}, the real part of the potential has been calculated using a folding model, and the
imaginary part of the potential has been determined by fitting to elastic-scattering data. Due to time limitations, especially in moving the detectors from the high-dispersion
focal plane to the medium-dispersion focal plane of the K600, it was not possible to take elastic-scattering data for this purpose. Instead, the Nolte, Machner and Bojowald optical-model potential was used. For this potential, the reduced radii are $r_R = 1.245$ fm
and $r_I = 1.570$ fm for the real and imaginary part of the potential, respectively.
Other parameters, such as the diffuseness and the depths of the potentials are
energy-dependent quantities, which are calculated separately for each entrance and exit channel.

For $^{24}$Mg, we employ the quadrupole deformation $\beta_2$ = 0.355 from Ref. \cite{VANDERBORG197931}. Using this deformation and the reduced radius of the real potential, we compute (with the codes {\sc belgen} and {\sc fermden} \cite{PC_MNH} which have been made available at \url{github.com/padsley/KVICodes}) the $B(E2)\uparrow$ using this deformation and the reduced radius of the real potential. This gives $B(E2)\uparrow = 0.0423$ $\mathrm{e}^2 \mathrm{b}^2$, which is in good agreement with the experimental value of $B(E2)\uparrow = 0.0432(19)$ $\mathrm{e}^2 \mathrm{b}^2$. Using the measured $B(E2)$ values  for $^{26}$Mg and $^{28}$Si, we obtain the quadrupole
deformations of $\beta_{2}$ = 0.295 and $\beta_{2}$ = -0.255, respectively. The signs of these deformations (prolate in $^{26}$Mg and oblate in $^{28}$Si) were chosen following the discussion in Sec. V-A. Note that the above parameters of the quadrupole deformation are much smaller than the absolute values for those from the NNDC database \cite{bnl} $\beta^{\rm exp}_2=0.613$, $0.484$, and $-0.412$ for $^{24,26}$Mg and $^{28}$Si. This is because the NNDC quadrupole deformation parameters are determined assuming a uniform charge distribution, while we use the Fermi distribution for the mass.

Since the radii of the real and imaginary parts of the potential are different,
we assumed that the deformation lengths for the real and imaginary parts of each
of the potentials are the same:
\begin{equation}
\beta_R R_R = \beta_I R_I
\label{beta_RI}
\end{equation}
 where $R_R = r_R A^{1/3}$,  $R_I = r_I A^{1/3}$, and $A$ is the mass number of the target  \cite{PhysRevC.23.2329}. Additionally, following Refs.
\cite{VANDERBORG1981243,harakeh2001giant}, we assume that the deformation lengths
of the potential and the mass distribution are identical, i.e.
\begin{equation}
\beta_R R_p = \beta_m R_m
\label{eq:beta_Rm}
\end{equation}
where the mass radius is $R_m = r_m A^{1/3}$ and $r_m$ is determined
from the reduced radius for the potential of Nolte, Machner and Bojowald
\cite{PhysRevC.36.1312}. Using the description by Satchler \cite{SATCHLER1987215},
the potential radius is $R_p = r_m (A^{1/3} + A_P^{1/3})$  where $A_P$ is
the mass of the projectile. The relation (\ref{eq:beta_Rm}) means
that the potential and mass distributions evolve self-consistently.

For monopole transitions, we used the {\sc formf} code \cite{PC_MNH,AngCorPackage}
to calculate the Satchler type-I  form factor \cite{SATCHLER1987215}.
For dipole transitions, we employed the form factors from Ref. \cite{PhysRevC.23.2329}.

For each excitation state, the $\beta^2_{R,\lambda}$ parameters were determined
by comparing the corresponding experimental and DWBA differential cross sections, see Eq. (\ref{eq:betaR-factor}). Then, using  Eqs. (\ref{beta_RI})
and (\ref{eq:beta_Rm}), the values $\beta_I$ and $\beta_m$ were obtained.

The percentage of the monopole $(\lambda=0)$ EWSR
exhausted by a given state is given by \cite{VANDERBORG1981243}:
\begin{equation}
 S_0 = \frac{\beta_{m,0}^2}{\beta_{M,0}^2},
\end{equation}
where $\beta^2_{m,0}$ is the monopole transition strength determined
from Eq. (\ref{eq:beta_Rm})  and
\begin{equation}
 \beta_{M,0}^2 = \frac{4\pi\hbar^2}{2 m A E_x \langle r^2\rangle}
\end{equation}
is the total transition strength for the state located at the excitation energy, $E_x$
and exhausting 100\% of the monopole EWSR \cite{VANDERBORG1981243}.
Here $m$ is the nucleon mass and $\langle r^2\rangle$
is calculated from the Fermi mass distribution using the {\sc fermden} code.

For dipole $(\lambda=1)$ transitions, the fraction of the EWSR
exhausted by a state is given by \cite{VANDERBORG1981243}:
\begin{equation}
 S_1 = \frac{\beta_{m,1}^2}{\beta_{M,1}^2},
\end{equation}
where $\beta^2_{m,1}$ is the dipole transition strength, again
from Eq. (\ref{eq:beta_Rm})  and
\begin{equation}
 \beta_{M,1}^2 = \frac{6\pi\hbar^2}{m A E_x}\frac{R_m^2}{11\langle r^4 \rangle
  - \frac{25}{3}\langle r^2 \rangle^2 - 10\epsilon \langle r^2 \rangle}
\end{equation}
is the total transition strength for the state lying at excitation energy,
$E_x$ and exhausting 100\% of the dipole EWSR
\cite{PhysRevC.23.2329}.
Here $\langle r^2 \rangle$ and $\langle r^4 \rangle$ are calculated from
the real part of the optical-model potential using {\sc fermden}, and $R_m$
is the half-density radius of the Fermi mass distribution. The parameter,
$\epsilon$, is generally small compared to the other quantities but is given by:
\begin{equation}
 \epsilon = \frac{\hbar^2}{3mA_T}\left( \frac{4}{E_2} + \frac{5}{E_0} \right),
\end{equation}
where $E_2 = 65 A^{-1/3}$ MeV is the centroid energy of the isoscalar
giant quadrupole resonance and $E_0 = 80 A^{-1/3}$ MeV is the centroid energy
of the isoscalar giant monopole resonance.

\bibliography{IS_March}
\end{document}